\documentclass[pre,twocolumn,showkeys,showpacs,groupaddress,preprintnumbers,floatfix,usenames,dvipsnames]{revtex4-1}

\usepackage{etex}
\usepackage{ifpdf}
\usepackage{hyperref}
\usepackage{dcolumn}
\usepackage{url}
\usepackage{amsmath}
\usepackage{amscd}
\usepackage{amsfonts}
\usepackage{amssymb}
\usepackage{amsthm}
\usepackage{bigints}
\usepackage{xfrac}
\usepackage{braket}
\usepackage{bm}   
\usepackage{bbm}
\usepackage{verbatim}
\usepackage{stmaryrd}
\usepackage{setspace}
\usepackage{overpic}
\usepackage{xcolor}

\usepackage{hyphenat}
\theoremstyle{plain}    
\theoremstyle{plain}    
\theoremstyle{plain}    
\theoremstyle{plain}    
\theoremstyle{plain}    
\theoremstyle{plain}    
\theoremstyle{plain}    
\theoremstyle{plain}    
\theoremstyle{plain}    
\theoremstyle{plain}    
\theoremstyle{plain}    
\theoremstyle{plain}    
\theoremstyle{plain}

\newcommand{\MeasAlphabet}  {\mathcal{A}}
\newcommand{\MeasSymbol}   { {X} }
\newcommand{\meassymbol}   { {x} }

\newcommand{\CausalState}   { \mathcal{S} }

\newcommand{\causalstate}   { \sigma }
\newcommand{\CausalStateSet}    { \boldsymbol{\CausalState} }
\newcommand{\AlternateState}    { \mathcal{R} }

\newcommand{\AlternateStateSet} { \boldsymbol{\AlternateState} }

\newcommand{\ProcessAlphabet}   {\MeasAlphabet}

\newcommand{\forward}{+}
\newcommand{\reverse}{-}
\newcommand{\forwardreverse}{\pm} 

\newcommand{\FutureCausalState} { {\CausalState}^{\forward} }

\newcommand{\PastCausalState}   { {\CausalState}^{\reverse} }

\newcommand{\lastindex}[2]{
  \edef\tempa{0}
  \edef\tempb{#2}
  \ifx\tempa\tempb
    \edef\tempc{#1}
  \else
    \edef\tempa{0}
    \edef\tempb{#1}
    \ifx\tempa\tempb
      \edef\tempc{#2}
    \else
      \edef\tempc{#1+#2}
    \fi
  \fi
  \tempc
}

\newcommand{\I}{\mathbf{I}}

\newcommand{\CSjoint}[1][,]{
   \edef\tempa{:}
   \edef\tempb{#1}
   \ifx\tempa\tempb
      \ensuremath{\FutureCausalState\!#1\PastCausalState}
   \else
      \ensuremath{\FutureCausalState#1\PastCausalState}
   \fi
}

\newif\ifpm
\edef\tempa{\forwardreverse}
\edef\tempb{\pm}
\ifx\tempa\tempb
   \pmfalse
\else
   \pmtrue
\fi

\renewcommand{\H}{\operatorname{H}}
\renewcommand{\I}{\operatorname{I}}

\colorlet {R_color}    {blue}
\colorlet {k_color}    {black!30!green}

\def\clap#1{\hbox to 0pt{\hss#1\hss}}

\begin{document}

\title{Fluctuations When Driving Between Nonequilibrium Steady States}

\author{Paul M. Riechers}
\email{pmriechers@ucdavis.edu}

\author{James P. Crutchfield}
\email{chaos@ucdavis.edu}

\affiliation{Complexity Sciences Center, Department of Physics\\
University of California at Davis, One Shields Avenue, Davis, CA 95616}

\date{\today}

\bibliographystyle{unsrt}

\begin{abstract}
Maintained by environmental fluxes, biological systems are thermodynamic
processes that operate far from equilibrium without detailed-balance dynamics.
Yet, they often exhibit well defined nonequilibrium steady states (NESSs). More
importantly, critical thermodynamic functionality arises directly from
transitions among their NESSs, driven by environmental switching. Here, we
identify constraints on excess thermodynamic quantities that ride above the
NESS housekeeping background. We do this by extending the Crooks fluctuation
theorem to transitions among NESSs, without invoking an unphysical dual
dynamics. This and corresponding integral fluctuation theorems determine how
much work must be expended when controlling systems maintained far from
equilibrium. This generalizes feedback control theory, showing that Maxwellian
Demons can leverage mesoscopic-state information to take advantage of the
excess energetics in NESS transitions. Altogether, these point to universal
thermodynamic laws that are immediately applicable to the accessible degrees of
freedom within the effective dynamic at any emergent level of hierarchical
organization. By way of illustration, this readily allows analyzing a
voltage-gated sodium ion channel whose molecular conformational dynamics play a
critical functional role in propagating action potentials in mammalian neuronal
membranes.
\end{abstract}


\preprint{Santa Fe Institute Working Paper 16-10-XXX}
\preprint{arxiv.org:1610.XXXX [cond-mat.stat-mech]}

\maketitle



\setstretch{1.1}

\newcommand{\Abet}{\ProcessAlphabet}
\newcommand{\MS}{\MeasSymbol}
\newcommand{\ms}{\meassymbol}
\newcommand{\SSet}{\CausalStateSet}
\newcommand{\St}{\CausalState}
\newcommand{\st}{\causalstate}
\newcommand{\MxSt}{\AlternateState}
\newcommand{\MxSSet}{\AlternateStateSet}
\newcommand{\mxst}{\eta}
\newcommand{\mxstw}[1]{\mxst_{#1}} 		
\newcommand{\StartMS}{\bra{\delta_\pi}}
\newcommand{\opGen}{A}         
\newcommand{\matHMM}{T}     
\newcommand{\matMSP}{W}     
\newcommand{\TT}{\mathcal{T}}

\newcommand{\HWA}{\ket{H(W^\Abet)}}
\newcommand{\Hmxst}{\ket{H[\mxst]}}

\newcommand{\corrbra}{\bra{\overline{\Abet}}}
\newcommand{\corrket}{\ket{\Abet}}

\newcommand{\Cmatrix}{\mathcal{C}}   
\newcommand{\LWwoutZero}{\Lambda_W^{\setminus 0}}

\newcommand{\pastx}{\smash{\overleftarrow x}}

\newcommand{\eeep}{\Omega}
\newcommand{\kB}{k_\text{B}}
\newcommand{\pastxt}{\smash{\overleftarrow x}_t}
\newcommand{\pastXt}{\smash{\overleftarrow X}_t}
\newcommand{\pastxtau}{\smash{\overleftarrow x}_\tau}
\newcommand{\pastXtau}{\smash{\overleftarrow X}_\tau}
\newcommand{\pastyt}{\smash{\overleftarrow y}_t}
\newcommand{\pastYt}{\smash{\overleftarrow Y}_t}
\newcommand{\pastytau}{\smash{\overleftarrow y}_\tau}
\newcommand{\pastYtau}{\smash{\overleftarrow Y}_\tau}
\newcommand{\pastX}{\smash{\overleftarrow X}}
\newcommand{\pastY}{\smash{\overleftarrow Y}}
\newcommand{\pastXY}{\smash{\overleftarrow {XY}}}
\newcommand{\pastxyt}{\smash{\overleftarrow {xy}}_t}
\newcommand{\pastXYt}{\smash{\overleftarrow {XY}}_t}
\newcommand{\MSPT}{\mathfrak{T}}
\newcommand{\T}{{\sf{T}}}
\newcommand{\It}{\mathcal{I}}
\newcommand{\Wxs}{W_\text{xs}}

\newcommand{\Qex}{Q_\text{ex}}
\newcommand{\Qhk}{Q_\text{hk}}
\newcommand{\Wex}{W_\text{ex}}
\newcommand{\Wdiss}{W_\text{diss}}

\newcommand{\pob}{\tfrac{\phi}{\beta}}

\newcommand{\xF}{\mathbf{x}}
\newcommand{\xR}{\mathbf{x}^\text{R}}
\newcommand{\sF}{\mathbf{s}}
\newcommand{\sR}{\mathbf{s}_{\leftarrow}^\text{R}}
\newcommand{\yF}{\mathbf{y}}
\newcommand{\yR}{\mathbf{y}^\text{R}}

\newcommand{\StartF}{\boldsymbol{\mu}_\text{F}}
\newcommand{\StartR}{\boldsymbol{\mu}_\text{R}}

\newcommand{\vx}{\vec{x}}
\newcommand{\vs}{\vec{s}}
\newcommand{\vy}{\vec{y}}
\newcommand{\vz}{\vec{z}}
\newcommand{\vxs}{\vec{xs}}
\newcommand{\vxz}{\vec{xz}}
\newcommand{\vxsz}{\vec{xsz}}
\newcommand{\vxy}{\vec{xy}}
\newcommand{\vxsy}{\vec{xsy}}

\newcommand{\vSF}{{\overrightarrow S}}
\newcommand{\vXF}{{\overrightarrow X}}
\newcommand{\vYF}{{\overrightarrow Y}}
\newcommand{\vsF}{{\overrightarrow s}}
\newcommand{\vxF}{{\overrightarrow x}}
\newcommand{\vyF}{{\overrightarrow y}}

\newcommand{\vxsF}{{\overrightarrow {xs}}}
\newcommand{\vXSF}{{\overrightarrow {XS}}}

\newcommand{\vSR}{{\overleftarrow S}}
\newcommand{\vXR}{{\overleftarrow X}}
\newcommand{\vYR}{{\overleftarrow Y}}
\newcommand{\vsR}{{\overleftarrow s}}
\newcommand{\vxR}{{\overleftarrow x}}
\newcommand{\vyR}{{\overleftarrow y}}

\newcommand{\RPD}{\rotatebox[origin=c]{180}{$\Upsilon$}}

\newcommand{\va}{v_\text{a}}
\newcommand{\vb}{v_\text{b}}

\section{Introduction}

The sun shines; ATP is abundant; power is supplied. These are the generous
settings in which we find many complex biological systems, buoyed steadily out
of equilibrium by energy fluxes in their environment. The resulting
steady-state dynamics exhibit various types of directionality, including
periodic oscillations and macroscopic thermodynamic functionality. These
behaviors contrast rather sharply with the deathly isotropy of equilibrium
detailed-balance dynamics---where fluxes are absent and state transition rates
depend only on relative asymptotic state-occupation probabilities.

Detailed balance and its implied dynamical reversibility, though, are common
tenets of equilibrium thermodynamics and statistical mechanics. They are
technically necessary when applying much of the associated theory relevant to
equilibrium and reversible (e.g., quasistatic) transitions between equilibrium
macrostates~\cite{Crooks11}. Detailed balance is even assumed by several modern
theorems that influence our understanding of the structure of fluctuations and
limitations on work performed far from equilibrium~\cite{Crooks98,Sagawa12}.
The natural world, though, is replete with systems that violate detailed
balance, such as small biological molecules constantly driven out of
equilibrium through interactions with their biochemical
environment~\cite{Wang98,Pole14}.

Far from happenstance and disruption, the probability currents through the
effective state-space of these nonequilibrium systems enable crucial
thermodynamic functionality~\cite{Land93,Qian01}. Even rare fluctuations play
an important functional role~\cite{Hors84,Lind04,Crut16a}. While constant
environmental pressure can drive a system into a nonequilibrium steady state
(NESS), complex biological systems are often driven farther---far from even any
NESS. Moreover, such system--environment dynamics involve feedback between
system and environment states. Although many believe these facilitate the
necessary complex processes that sustain life, their very nature seems to
preclude most, if not all, hope of a universal theoretical framework for
quantitative predictions. To ameliorate the roadblock, we present a consistent
thermodynamics that is not only descriptive, but constructive, tractable, and
predictive, even when irreversible dynamics transition between NESSs.

Beyond laying out the structure of fluctuations during NESS transitions, this
thermodynamics sets the stage to understand how one level of organization gives
way to another. In particular, using it a sequel renormalizes nonequilibrium
housekeeping background to show how to maintain a hierarchy of steady-state
dynamics. Said simply, at each level of hierarchical organization, controllable
degrees of freedom are subject to universal thermodynamic laws that tie their
fluctuations and functionality to dissipation at lower levels.


\subsection{Results}

Nonequilibrium thermodynamics progressed markedly over the last two decades on
at least two fronts. First, by taking the `dynamics' in `thermodynamics'
seriously, fluctuation theorems (FTs) transformed previous inequalities, such
as the classical Second Law of Thermodynamics, into subsuming equalities that
exactly express the distribution of thermodynamic variations.  (These have been
derived by many authors now in a wide range of physical settings; see, e.g.,
Refs.~\cite{Seifert12} and \cite{Spin13} for lucid reviews.) Second,
steady-state thermodynamics (SST) showed that NESSs play a role in
nonequilibrium analogous to that of equilibrium macrostates in equilibrium. In
this view heat decomposes into the \emph{housekeeping heat} $\Qhk$ needed to
sustain NESSs and the \emph{excess heat} $\Qex$ dissipated in transitions
between them \cite{Oono98,Hatano01,Trepagnier04}. Bolstering SST, recent
efforts generalized the Clausius inequality (describing excess heat produced
beyond the change in system entropy) to smoothly driven transitions between
NESSs~\cite{Mandal15}. Taken together, these results established an integral
fluctuation theorem for the excess work in NESS transitions and, consequently,
a generalized Second Law for excess entropy produced beyond housekeeping during
driven NESS transitions.

The following extends SST by introducing several new FTs, highlighting
correspondences between nonequilibrium and equilibrium relations. First, we
provide detailed (i.e., nonintegrated) fluctuation theorems, rather than
integral fluctuation theorems for driven NESS transitions. (Integral FTs
follow directly, in any case.) This constrains distributions of excess work $\Wex$
exerted when controlling nonequilibrium systems. Second, we jointly bound
housekeeping and excess work distributions. For example, for time-symmetric
driving we show that the joint probability of excess work and housekeeping heat
respect the strong constraint:
\begin{align*}
\frac{\Pr(\Wex, \Qhk)}{\Pr(-\Wex, -\Qhk)} = e^{\beta \Wex} e^{\beta \Qhk}
  ~.
\end{align*}
When the transitions are nonequilibrium excursions between \emph{equilibrium}
steady states, this reduces to the Crooks FT~\cite{Crooks99}. Third, we derive
the detailed FTs for entropy production even when temperature varies in space
and time. They are expressed in terms of excess environmental entropy
production $\eeep$ and irreversibility $\Psi$, even when the irreversibility is
``housekeeping'' not strictly associated with heat. Finally, we quantify a
system's net \emph{path irreversibility} $\Psi$ with the accumulated violation
of detailed balance in the effective dynamic. In the isothermal setting, for
example, the irreversibility is the housekeeping heat, maintaining the system
in its nonequilibrium dynamic: $\Psi = \beta \Qhk$. Importantly, we can
determine the minimum housekeeping heat without appealing to the system's
Hamiltonian.

Extending SST in this way reveals universal constraints on excess thermodynamic
quantities---effective energies accessible above the housekeeping background.
Looking forward, this allows one to analyze nondetailed-balanced stochastic
dynamics---and thus contributes an understanding of the role of hierarchy---in
the thermodynamics of replication~\cite{England13} and the thermodynamics of
learning \cite{Crut92c}. Moreover, this identifies how complex, possibly
intelligent, thermodynamic systems leverage (designed or intrinsic)
irreversibility in their own state-space to harness energy from structured
environments.

\subsection{Synopsis}

Section~\ref{sec:Setup} sets up our approach, introducing notation, discussing
input-dependent system dynamics, and establishing fundamental relationships
among nonequilibrium thermodynamic quantities. Section~\ref{sec:WorkHeatEtc}
introduces excess heat and excess work in analogy to classical heat and work.
Ultimately though, the related excess environmental entropy production $\eeep$
discussed in \S~\ref{sec:EEEP} generalizes these to the case of temperature
inhomogeneity over spacetime. Section~\ref{sec:PathEntropies} demonstrates that
path entropies are the fundamental objects of nonequilibrium thermodynamics. In
steady state, unaveraged path entropies reduce to the steady-state surprisal
$\phi$. Deviations from the asymptotic surprisal contribute to a
nonsteady-state additional free energy. All of these quantities play a central
role in the subsequent development.

Before delving into irreversibility, though, we first address what is meant by
reversibility. Therefore, \S~\ref{sec:DetailedBalance} and
\S~\ref{sec:MicroReversibility} discuss detailed balance, microscopic
reversibility, and the close relationship between them.
Section~\ref{sec:Irreversibility} then introduces path dependence and
reverse-path dependence and explains how together they yield a system's
irreversibility $\Psi$.

With this laid out, \S~\ref{sec:GenDetailedFT} and \S~\ref{sec:GenCFT} derive
the detailed FTs in terms of excess environmental entropy production $\eeep$
and irreversibility $\Psi$. One sees that in the isothermal setting $\Psi =
\beta \Qhk$ and the excess entropy production is directly related to the excess
work. This allows \S~\ref{sec:SST_FTs} to explain how these results extend SST.

Sections~\ref{sec:IntegralFTs} and \ref{sec:FeedbackFTs} finish our
investigation of NESS FTs by deriving several integral FTs. This, in effect,
extends feedback control, as developed in Refs. \cite{Sagawa12} and
\cite{Horowitz10}, to SST. We note that such environmental feedback is
intrinsic to natural systems.

For concreteness, \S~\ref{sec:IonChannel} analyzes a simple but biologically
important prototype system: voltage-gated sodium ion channels. These are
complex macromolecules that violate detailed balance in order to perform
critical biological functioning far from equilibrium. Finally, appendices
discuss non-Markovian dynamics and comment on the bounds provided by
integral fluctuation theorems for auxiliary variables.

\section{Driven Stochastic Dynamics}
\label{sec:Setup}

We consider a classical system---the \emph{system under study}---with
time-dependent driving via environmentally determined parameters; e.g.,
time-dependent temperature, voltage, and piston position. Hence, the
environmental control input $X_t$ at time $t$, taking on values $x_t \in
\mathcal{X}$, will typically be a vector object. The system under study is
assumed to have a countable set $\SSet$ of states. The random variable $\St_t$
for the state at time $t$ takes on values $s_t \in \SSet$. We assume that the
environment's control value (current \emph{input}) $x$ and the system's
physical state (current \emph{state}) $s$ are sufficient to determine the
system's net effective energy---the \emph{nonequilibrium potential}
$\phi(x,s)$. Even with constant environmental input, the system dynamic need
not be detailed balance.

\subsection{Stochastic mesoscopic dynamics and induced state-distributions}

We assume the current environmental input $x$ determines the instantaneous
stochastic transition dynamic over the system's observable mesoscopic states.
However, that input can itself depend arbitrarily on all previous input and
state history. That is, we assume that the $\SSet$-to-$\SSet$ transitions are
instantaneously Markovian given the input. Over time, though, different inputs
induce different Markov chains over system states.

Note that the Markov assumption is common, although often implicit, and we
follow this here to isolate the novel implications of nondetailed-balance
dynamics. Nevertheless, the results generalize to infinite Markov order by
modeling system states as the observable output of many-to-one mappings of
latent states of an input-controllable hidden Markov chain.
Appendix~\ref{sec:HMMs} details this generalization.

We do not restrict the environment's driving process, allowing arbitrary
non-Markovity, feedback, and nonstationarity. Thus, the joint
system-environment dynamic can be non-Markovian even if the instantaneous
system dynamic is. Such a setup is quite general, and so the results to follow
extend others known for SST. We also follow stochastic thermodynamics in the
use of (arbitrarily small) discrete-time steps. Nevertheless, it is usually
easy to take the continuous-time limit. As, in fact, we do in the example
at the end.

Hence, the Markovian dynamic is described by a (possibly infinite) set of
input-conditioned transition matrices over the state set $\SSet$: $\{
\T^{(\SSet \to \SSet | x)} \}_{x \in \mathcal{X}}$, where $\T^{(\SSet \to \SSet
| x)}_{i,j} = \Pr(\St_{t} = s^j | \St_{t-1} = s^i , X_t = x)$ is the
probability that the system is in state $s^j$ at time $t$ given that the system
was in state $s^i$ at time $t-1$ and the instantaneous environmental input
controlling the system was $x$.

The Perron--Frobenius theorem guarantees that there is a stationary
distribution $\boldsymbol{\pi_x}$ over states associated with each fixed input
$x$. These are the state distributions associated with the system's
nonequilibrium steady states (NESSs). For simplicity, and unless otherwise
stated, we assume that a fixed input $x$ eventually induces a unique NESS.

We denote distributions over the system states as bold Greek symbols; such as
$\boldsymbol{\mu}$. We denote the state random variable $\St$ being
distributed according to $\boldsymbol{\mu}$ via $\St \sim \boldsymbol{\mu}$.
It will often be convenient to cast $\boldsymbol{\mu}$ as a row-vector, in
which case it appears as the bra $\bra{\boldsymbol{\mu}}$.  Putting this
altogether, a sequence of driving inputs updates the state distribution as
follows:
\begin{align*}
\bra{\boldsymbol{\mu}_{t+n}} 
&= \bra{\boldsymbol{\mu}_{t}} \T^{(\SSet \to \SSet | x_{t:t+n} )} \\
&= \bra{\boldsymbol{\mu}_{t}} \T^{(\SSet \to \SSet | x_{t} )} \T^{(\SSet \to \SSet | x_{t+1} )} \cdots \T^{(\SSet \to \SSet | x_{t+n-1} )} ~.
\end{align*}
(Time indexing here and throughout is denoted by subscripts $t\!:\!t'$ that are
left-inclusive and right-exclusive.) An infinite driving history $\vxR = \ldots
x_{-2} x_{-1}$ induces a distribution $\boldsymbol{\mu}(\vxR)$ over the state
space. The so-called \emph{steady-state distribution} associated with the
environmental drive value $x$, induced by tireless repetition of $x$, is: 
\begin{align*}
\bra{\boldsymbol{\pi}_x}
  = \lim_{n \to \infty}
  \bra{\boldsymbol{\mu}_0} \left( \T^{(\SSet \to \SSet | x)} \right)^n
  ~.
\end{align*}
Usefully, $\boldsymbol{\pi}_x$ can also be found as the left eigenvector of
$\T^{(\SSet \to \SSet | x)}$ associated with the eigenvalue of unity
\footnote{We ignore nonergodicity to simplify the development. The approach,
though, handles nonergodicity just as well. However, distracting nuances arise
that we do not wish to dwell on. For example, if the Markov chain has more than
one attracting component for a particular $x$, then $\boldsymbol{\pi_x}$ is not
unique, but can be constructed as any one of infinitely many
probability-normalized linear superpositions of left eigenvectors of
$\T^{(\SSet \to \SSet | x)}$ associated with the eigenvalue of unity.}:
\begin{align}
\bra{\boldsymbol{\pi}_x} = \bra{\boldsymbol{\pi_x}} \T^{(\SSet \to \SSet | x)} ~.
\label{eq:eigenpix1}
\end{align}

The assumption that observable state-to-state transitions are instantaneously
Markovian allows the state distribution $\boldsymbol{\mu}$ to summarize the
causal relevance of the entire driving history $\vxR$.

\section{Energetics and Entropies}

For later comparison, we recount the basics of a statistical mechanics
description of the thermodynamics of a system exchanging energy with a large
environment, imposing fixed constraints indexed as $x$. The many-body
Hamiltonian $\mathcal{H}(x)$ has energy eigenvalues $\{ E(x,s) \}$, where $s$
indexes the energy eigenstates. The canonical distribution is $\pi_x(s) = e^{-
\beta [E(x,s) - F_\text{eq}(x)] }$, at fixed $x$. This distribution is the
equilibrium steady ``state'' associated with $x$, where $\beta^{-1} \equiv \kB
T$, $T$ is the temperature of the macroscopic environment surrounding the
system, and $F_\text{eq}(x)$ is the associated equilibrium free energy.

\subsection{Work, heat, and their excesses}
\label{sec:WorkHeatEtc}

\emph{Work} $W$ is environmentally driven energy change. Within one time-step
it is given by~\footnote{We start in a discrete-time setup, but later translate
to continuous time}:
\begin{align*}
W[x_{n-1} \to x_n ; s_{n-1}] = E(x_n, s_{n-1}) - E(x_{n-1}, s_{n-1}) ~.
\end{align*}
\emph{Heat} $Q$ is the change in system energy due to its internal response to
the environmental drive; e.g., a molecule's change in conformation. Within one
time-step the heat is:
\begin{align*}
Q[x_{n}  ; s_{n-1} \to s_n] = E(x_n, s_{n}) - E(x_{n}, s_{n-1}) ~.
\end{align*}
Over the course of driving the system from $t = 0$ to $t = N \Delta t = \tau$,
the net energy change is then:
\begin{align*}
\Delta E & = E(x_N, s_{N}) - E(x_{0}, s_{0}) \\
         & = W + Q ~,
\end{align*}
where the net work and net heat are:
\begin{align*}
W = \sum_{n = 1}^N  W[x_{n-1} \to x_n ; s_{n-1}]  
\end{align*}
and:
\begin{align*}
Q = \sum_{n = 1}^N Q[x_{n}  ; s_{n-1} \to s_n]
  ~,
\end{align*}
respectively. Here, and later on, $\Delta$ applied to a quantity refers to its
change over one time step, where the step is given by context.

When the system strongly couples to a substrate with uncontrolled energy fluxes,
steady-state dynamics are often established far from equilibrium, even when
environmental parameters are held fixed. That is, for fixed driving $\dots
xxxxx \dots$, the system settles down to a NESS with a distribution over
observable system states given by the \emph{nonequilibrium potential}
$\phi(x,s)$:
\begin{align}
\pi_x(s) = e^{- \phi(x,s) }
  ~.
\label{eq:NEPdef}
\end{align}
In this, $\phi(x,s)$ plays a role roughly analogous to energy eigenvalues.
Thus, the thermodynamics of accessible energetics---the excess heat generated
and work irretrievably performed in driving between NESSs---follows analogously
to its equilibrium counterpart. This is complementary to recent SST
studies~\cite{Hatano01,Trepagnier04,Espo10,Mandal15}.

If steady-state free energies $F_\text{ss}(x)$ and \emph{effective} energies $E_\text{eff}(x,s)$ could be uniquely 
(and usefully) defined, then the nonequilibrium potential would be:
\begin{align*}
\phi(x,s) = \beta [ E_\text{eff}(x,s) - F_\text{ss}(x) ]
  ~.
\end{align*}
However, the assignment of steady-state free energies is problematic.
Nevertheless, $\phi(x,s)$ retains meaning since it quantifies the steady-state
\emph{surprisal} of observing state $s$:
\begin{align*}
\phi(x,s) = - \ln \pi_x(s)
  ~.
\end{align*}
The surprisal is Shannon's \emph{self-information} \cite{Cove06a}---the
unaveraged individual-event entropy measuring how surprising a specific event
is. Intuitively, we must do work to make otherwise unlikely things happen.

SST's \emph{excess} work and heat can be defined via changes in steady-state
surprisal $\phi$, analogous to how equilibrium quantities are in terms of
energy changes. For clarity, we temporarily restrict ourselves to the
isothermal setting, but we can easily adapt to time-varying temperatures.

\emph{Excess work} $\Wex$ is environmentally driven change in nonequilibrium
potential:
\begin{align*}
\Wex [x_{n-1} & \to x_n ; s_{n-1}] \\
  & = \beta^{-1} [ \phi(x_n, s_{n-1}) - \phi(x_{n-1}, s_{n-1}) ]
  ~,
\end{align*}
over one time-step. \emph{Excess heat} $\Qex$ is the change in nonequilibrium
potential due to the system's response:
\begin{align*}
\Qex [x_{n}  ; s_{n-1} \to s_n] = \beta^{-1} [\phi (x_n, s_{n}) - \phi (x_{n}, s_{n-1}) ] ~,
\end{align*}
over one time-step. When driving from $t = 0$ to $t = N
\Delta t = \tau$, the net change in nonequilibrium potential is:
\begin{align}
\Delta \phi & = \phi (x_N, s_{N}) - \phi (x_{0}, s_{0}) \nonumber \\
            & = \beta (\Wex + \Qex ) \nonumber \\
            & = - \ln \frac{\pi_{x_N}(s_N)}{\pi_{x_0}(s_0)}
  ~,
\label{eq:NoneqPotential}
\end{align}
where the net excess work and net excess heat are:
\begin{align}
\Wex = \sum_{n = 1}^N  \Wex [x_{n-1} \to x_n ; s_{n-1}]  
\label{eq:WexDefined}
\end{align}
and:
\begin{align}
\Qex = \sum_{n = 1}^N \Qex [x_{n}  ; s_{n-1} \to s_n]
  ~,
\label{eq:QexDefined}
\end{align}
respectively.

This approach to excess heat $\Qex$ coincides with SST's definition and reduces
to total heat in equilibrium transitions. Importantly, it follows as closely as
possible the equilibrium approach to total heat $Q$ outlined above and deviates
from the typical starting point: $\Qex \equiv Q - \Qhk$, where $\Qhk$ is the
so-called \emph{housekeeping heat}. In contrast, excess
work $\Wex$ does \emph{not} reduce to the total work in equilibrium
transitions. Rather, $\Wex$ goes over to $W - \Delta F_\text{eq}$, if the
steady states are near equilibrium. And, this fortuitously coincides with its
previous narrower use in describing transitions atop equilibrium steady
states---the work exerted beyond the change in free energy~\cite{Still12}.

The excess heat $\Qex$ can be interpreted as the heat dissipated during
transitions between NESSs. Similarly, the excess work $\Wex$ can be interpreted
as the work that \emph{would} be dissipated if the system is allowed to relax
back to a NESS. The difference between excess work $\Wex$ and \emph{dissipated
work}, denoted $\Wdiss$, depends on a notion of excess nonequilibrium free
energy, discussed shortly.

This framing leads us to see that heat is how small, possibly intelligent,
systems store and transform energy via their \emph{own agency}. This stance
also moves us away from any unjustified biases that heat is necessarily
wasteful. For example, an increase in heat may indicate that a system has
harvested energy, and the emission of heat may indicate an intrinsic
computation \cite{Crut88a} in the system's state-space. The \emph{efficiency}
of the tradeoff---spending stored energy to achieve some utility---then comes
into question. It is inefficiency in this sense that is by its nature wasteful.

\subsection{Excess environmental entropy production}
\label{sec:EEEP}

In isothermal transitions between equilibrium steady states, 
the environmental entropy production is~\cite{Crooks99}: 
\begin{align*} 
\eeep_\text{eq} & = \beta (W - \Delta F_\text{eq}) \\
    & = -\beta Q - \ln \frac{\pi_{x_N}(s_N)}{\pi_{x_0}(s_0)}
	~.
\end{align*}
This extends to SST by defining the \emph{excess environmental entropy production}:
\begin{align} 
\eeep & = \beta \Wex \nonumber \\
      & = -\beta \Qex - \ln \frac{\pi_{x_N}(s_N)}{\pi_{x_0}(s_0)}
	  ~,
\label{eq:EEEPdef}
\end{align}
This has also been referred to as the ``nonadiabatic component of entropy
production''~\cite{Espo07, Espo10,Bagc12a,Mandal15}. Note that $-\ln\left(
\pi_{x_N}(s_N) / \pi_{x_0}(s_0) \right) = \Delta \phi$, recovering Eq.
(\ref{eq:NoneqPotential})'s change in nonequilibrium potential $\phi$.

Recalling the definitions of $\Qex$ in terms of steady state surprisals and
$\phi(x,s) = - \ln \pi_x(s)$, we see that:
\begin{align}
e^{-\beta \Qex}
  & = e^{- \sum_{n = 1}^N
    \left[ \phi(x_n, s_{n}) - \phi(x_{n}, s_{n-1}) \right] } \nonumber \\ 
  & = \prod_{n=1}^N \frac{\pi_{x_n}(s_{n})}{\pi_{x_n}(s_{n-1})}
  ~. 
\label{eq:embqAsProduct}
\end{align}
And so, Eq. (\ref{eq:EEEPdef}) gives:
\begin{align}
e^{\eeep(x_{0:N+1}, s_{0:N})}
  & = \frac{\pi_{x_0}(s_0)}{\pi_{x_N}(s_N)} \prod_{n=1}^N 
  	\frac{\pi_{x_n}(s_{n})}{\pi_{x_n}(s_{n-1})} \nonumber \\
  & = \prod_{n = 0}^{N-1} \frac{\pi_{x_n}(s_{n})}{\pi_{x_{n+1}}(s_{n})}
  ~.
\label{eq:ewAsProduct}
\end{align}

If temperature varies, then the above still holds if we replace the equilibrium 
probabilities with the temperature-dependent equilibrium probabilities.
Thus, to go beyond the isothermal setting, we use Eq.~\eqref{eq:ewAsProduct} as
the defining relationship for the excess environmental entropy production
$\eeep$. If temperature is spatially homogeneous, then it is equivalent to:
\begin{align*}
\eeep = \Delta \phi - \tfrac{1}{\kB} \int \frac{ \delta \Qex }{T} ~.
\end{align*}
However, spatially inhomogeneous temperatures can also be addressed by folding
temperature dependence into the environmental input $x$.

We return to these expressions and explore their role in generalized
fluctuation theorems once we develop the necessary quantitative notions of
irreversibility.

\subsection{Path entropies}
\label{sec:PathEntropies}

In steady state, the system state probability distribution has a Boltzmann
exponential dependence on the effective energies. Naturally, out of steady
state the distribution is something different. There is a nonsteady-state free
energy associated with this out-of-steady-state distribution, since the system
can do work (or computations) at the cost of relaxing the distribution.

Nonsteady-state free energies are controlled by path entropies, which come in
several varieties. Here, we are especially interested in the controllable
unaveraged state surprisals induced by the driving path $\vxR$:
\begin{align}
h^{(s | x_{-\infty:t+1})} = - \ln \Pr(\St_t = s | x_{-\infty:t+1}) ~.
\label{eq:PathEnt}
\end{align}
Since a semi-infinite history induces a particular distribution over system states, this can be usefully recast in terms of the 
initial distribution $\boldsymbol{\mu}_0$ induced by the path $x_{-\infty:1}$
and the driving history $x_{1:t+1}$ since then:
\begin{align}
h^{(s | \boldsymbol{\mu}_0 , x_{1:t+1} ) } 
  & = - \ln \Pr(\St_t = s | \St_0 \sim \boldsymbol{\mu}_0 , x_{1:t+1}) \\
  & = -  \ln \bra{\boldsymbol{\mu}_0} \T^{(\SSet \to \SSet | x_{1:t+1})}  \ket{s} ~, \nonumber
\end{align}
where $\Pr(\St_t = s | \St_0 \sim \boldsymbol{\mu}_0 , x_{1:t+1})$ is the
probability that the state is $s$ at time $t$, under the measure induced when
the initial state $\St_0 \sim \boldsymbol{\mu}_0$ (distributed according to
$\boldsymbol{\mu}_0$)~\footnote{To be more precise, we write $\Pr(\St_t = s |
\St_0 \sim \boldsymbol{\mu}_0 , x_{1:t+1})$ as $\Pr_{\St_0 \sim
\boldsymbol{\mu}_0}(\St_t = s | x_{1:t+1})$, since the probability is not
\emph{conditioned} on $\boldsymbol{\mu}_0$---a probability measure for subsequent
state sequences. Here, we simply gloss over this nuance, later adopting the
shorthand: $\Pr(\St_t = s | \boldsymbol{\mu}_0 , x_{1:t+1})$.} and given the
driving history $x_{1:t+1} = x_1 \dots x_t$ since the initial time.

Alternatively, consider the distribution $\boldsymbol{\mu}$ induced from a
start distribution by the driving history since the start. Then the
path-induced state-surprisal can be expressed simply in terms of the
\emph{present} environmental-history-induced distribution over system states
and the candidate state $s$:
\begin{align}
h^{(s | \boldsymbol{\mu} ) } 
  & = - \ln \Pr(\St_t = s | \St_{t} \sim \boldsymbol{\mu} ) \\
  & = - \ln \braket{\boldsymbol{\mu} | s}
  \nonumber
  ~.
\end{align}
Thermodynamic units of entropy are recovered by multiplying the Shannon-like
path surprisals by Boltzmann's constant: $\mathfrak{s} = \kB h$.

Averaging the path-induced state-surprisal over states gives a genuine input-conditioned Shannon entropy:
\begin{align}
\braket{h^{(s_t | \pastxt)} }_{\Pr(s_t | \pastxt)} 
  & = - \Bigl\langle \ln \Pr(s_t | \pastxt )
  \Bigr\rangle_{\Pr( s_t | \pastxt )} \nonumber \\
  & = -  \sum_{s_t}  \Pr(s_t | \pastxt)
  \ln \Pr(s_t | \pastxt ) \nonumber \\
  & =  \H [\St_t | \pastXt = \pastxt ]  ~,
\label{eq:HSx}
\end{align}
where $\H[\cdot | \cdot]$ is the conditional Shannon entropy in units of nats.

It follows directly that the state-averaged path entropy $\kB \H[\St_t |
\pastxt ]$ is an extension of the system's steady-state nonequilibrium entropy
$S_\text{ss}$. In steady-state, the state-averaged path entropy reduces to:
\begin{align}
\kB \H[\St_t | \pastXt = \dots xxx ]
  & = - \kB \H[\St_t | \St_t \sim \boldsymbol{\pi_x} ]  \nonumber \\ 
  & = - \kB \sum_{s \in \SSet} \pi_x(s) \ln \pi_x(s) \label{eq:Sss} \\
  & = S_\text{ss}(x) \nonumber
  ~.
\end{align}
The system steady-state nonequilibrium entropy $S_\text{ss}$ has been discussed
as a fundamental entity in SST; e.g., see Refs. \cite{Hatano01} and
\cite{Mandal15}. However, Eq.~\eqref{eq:HSx} ($\times \kB$) gives the
appropriate extension for the thermodynamic entropy of a nonequilibrium system
that is \emph{not in steady state}. Rather, it is the entropy over system
states given the entire history of environmental driving.

When $\SSet$ is the set of microstates, rather than, say, observable mesoscopic
states, the unaveraged nonequilibrium free energy $F$ enjoys the familiar
relationship between energy $E$ and (path) entropy $\mathfrak{s}$:
\begin{align}
& F^{(s_t | x_{-\infty:t+1} ) } 
  \equiv E(x_t, s_t) - T \mathfrak{s}^{(s_t | x_{-\infty:t+1} ) } 
	\label{eq:FETS} \\
  & \quad = F_\text{eq}(x_t) + \beta^{-1}
  	\ln \frac{\Pr(s_t | x_{-\infty:t+1} )}{ \pi_{x_t}(s_t ) }
  ~.
\label{eq:FDKL}
\end{align}
Or, averaging over states:
\begin{align}
\mathcal{F}(t) & = U(t) -  \beta^{-1} \H[\St_t | x_{-\infty:t+1} ] 
	\label{eq:FUTS}  \\
  & = F_\text{eq}(x_t) +  \beta^{-1} D_\text{KL}
  \bigl( \Pr(\St_t | x_{-\infty:t+1} ) \, || \, \boldsymbol{\pi}_{x_t} \bigr)
  \nonumber
  ~,
\end{align}
where $\mathcal{F}(t)$ is the expected instantaneous nonequilibrium free
energy, $U(t)$ is the expected instantaneous thermal energy, and $D_\text{KL}(
\cdot || \cdot )$ is the Kullback--Leibler divergence~\cite{Cove06a}.
Recognizing $\kB \H[\St_t | x_{-\infty:t+1} ]$ as the natural extension of a
system's thermodynamic entropy, Eq.~\eqref{eq:FUTS} is familiar from
equilibrium thermodynamics, but it is now applicable arbitrarily far from
equilibrium and at any time $t$ using the instantaneous temperature. This is not
the first statement of such a generalized relationship; compare, e.g.,
Refs.~\cite{Gaveau97, Sivak12}. In equilibrium, the expected value of the path
entropy (using microstates) reduces to \emph{the} equilibrium entropy of a
system.

In the setting of effective states and NESS surprisals, we can no longer
directly use Eq.~(\ref{eq:FETS}). Nevertheless, by analogy with
Eq.~\eqref{eq:FDKL}, we can still identify the \emph{nonsteady-state addition
$\gamma(\cdot|\cdot)$ to free energy} as:
\begin{align*}
\beta^{-1} \gamma(s | \boldsymbol{\mu} , x)
  \equiv \beta^{-1}
  \ln
  \frac{\Pr(\St_t = s | \St_{t-1} \sim \boldsymbol{\mu}, X_{t} = x )}
  { \pi_{x}(s ) }
  ~.
\end{align*}
Expressed differently, it is:
\begin{align*}
\gamma(s | \boldsymbol{\mu} , x) 
&= h^{(s | \boldsymbol{\pi_x} ) } - h^{(s | \boldsymbol{\mu} , x ) } \\
&= \phi(x,s) - h^{(s | \boldsymbol{\mu} , x ) }
~.
\end{align*}
Averaging over states this becomes the Kullback--Leibler divergence between
nonsteady state and steady state distributions:
\begin{align*}
\braket{\gamma(s | \boldsymbol{\mu} , x) }
  =  D_\text{KL} \bigl[ \Pr(\St_t  | \St_{t-1} \sim \boldsymbol{\mu} ,  X_{t} = x ) \, || \, \boldsymbol{\pi}_{x} \bigr]
  ~,
\end{align*}
which is nonnegative.

Identifying the nonsteady-state contribution to the free energy allows us to introduce the \emph{dissipated work}:
\begin{align*}
W_\text{diss} \equiv \Wex - \beta^{-1} \Delta \gamma ~,
\end{align*}
to account for the fact that excess work is not fully dissipated until the
distribution relaxes back to steady state $\boldsymbol{\pi}_{x}$. An important
consequence is that the excess work dissipated can be reclaimed by a subsequent
``fluctuation'' with $\Wex < 0$ in the midst of a driven nonequilibrium
excursion.

The role of the nonsteady-state contribution to free energy will be apparent in
the FTs to come shortly. This generalizes similar FTs that are restricted to
starting and possibly ending in a steady state $\boldsymbol{\pi}_{x}$. The
generalization here is key to analyzing complex systems, since many simply
cannot be initiated in a steady state without losing their essential character.

\section{Irreversibility}

To emphasize, the preceding did not reference and does not require detailed
balance. However, to ground the coming development, we need to describe the
roles of reversibility, detailed balance, and their violations. At a minimum,
this is due to most FTs assuming reversibility of the effective dynamic over
states. Having established the necessary concepts and giving a measure of the
irreversibility of the effective dynamic, we finally move on to FTs for
nondetailed balanced processes.

\subsection{Detailed balance}
\label{sec:DetailedBalance}

Transitioning from state $a$ to state $b$, say, invoking detailed balance
assumes that:
\begin{align*}
\frac{\Pr(\St_n = a | \St_{n-1} = b, X_n = x)}{\Pr(\St_n = b | \St_{n-1} = a, X_n = x)}
= \frac{\pi_{x}(a)}{\pi_{x}(b)} ~.
\end{align*}
Though, we do \emph{not} assume detailed balance over the states considered
here, we refer to it occasionally. For example, assuming detailed balance,
microscopic reversibility and the standard Crooks fluctuation theorem (CFT)
follow almost immediately.

In contrast, complex systems sustained out of equilibrium by an active
substrate generically evolve via nondetailed-balance dynamics. To wit, many
examples of nondetailed-balance dynamics are exhibited by chemical kinetics in
biological systems~\cite{Qian05, Liep07a, Liep07b}. We conclude with a
thorough-going thermodynamic analysis of one neurobiological example.

\subsection{Microscopic reversibility}
\label{sec:MicroReversibility}

Consider a particular realization of interleaved environmental-input sequence
$\ldots x^1 x^2 \cdots x^{N-1} \ldots$ and system-state sequence $\ldots s^1 s^2
\cdots s^{N-1} \ldots$:
\begin{align*}
\textcolor{gray}{x^0} 	\quad \,  \overbrace{x^1 \quad \, x^2 \quad \quad \dots   	\quad \quad  x^{N-1} }^{\xF , \, \xR} \quad  \textcolor{gray}{x^N} \quad   \\
 \quad  \underbrace{ s^0 \quad \, \,  \underbrace{ s^1  \quad \, \, s^2  \quad \;  \quad  s^{N-2} \quad  s^{N-1} }_{\sF} \!\!\!\!\!\!\! \!\!\!\!\!\!\! \!\!\!\!\!\!\! }_{\sR} \qquad \qquad  \textcolor{gray}{s^N}
\end{align*}

There are several length-$(N-1$) subsequences in play here, including the
forward trajectory $\xF = x^1 x^2 \cdots x^{N-2} x^{N-1}$ of the environmental
driving and the forward trajectory $\sF = s^1 s^2 \dots s^{N-2} s^{N-1}$ of the
state sequence. Furthermore, let $\xR = x^{N-1} x^{N-2} \dots x^2 x^1$ be the
time-reversal of the environmental driving $\xF$ and $\sR = s^{N-2} s^{N-3}
\dots s^1 s^0$ the time-reversal of the time-shifted state sequence $\sF$.

For example, if $\mathcal{X} = \{ 0, 1\}$ and $\SSet = \{ a, b, c \}$, then
$\xF$ may be the sequence $00101110 \dots 11000010$ and $\sF$ the sequence
$acaaaaba \dots abaccabc$. Then $\xR$ is the sequence $01000011 \dots
01110100$. Taking the time reversal of the state sequence, we have $cbaccaba
\dots abaaaaca$. However, since $\sR$ is also time-shifted by one time-step, we
must drop the first $c$ and append another symbol, say $a$. Then $\sR$ is the
sequence $baccaba \dots abaaaacaa$.

Let $Q_\text{F}$ be the excess heat of the joint forward sequences
$\xF$ and $s^0 \sF$,
according to Eq.~\eqref{eq:QexDefined}.
By definition, a system--environment effective dynamic is \emph{microscopically reversible} if:
\begin{align*}
\frac{\Pr(\St_{1:N} = \sF | \St_0 = s^0, X_{1:N} = \xF)}
{\Pr(\St_{1:N} = \sR  | \St_0 = s^{N-1}, X_{1:N} = \xR)}
& = e^{-\beta Q_\text{F}}
  ~,
\end{align*} 
for any
$s^0 \in \SSet$, $\sF \in \SSet^{N-1}$, and $\xF \in \mathcal{X}^{N-1}$.
As a useful visual aid, we can re-express this as:
\begin{align*}
\frac{\Pr(s^0 \xrightarrow{x^1} s^1 \cdots s^{N-2} \xrightarrow{x^{N-1}} s^{N-1} | s^0, \xF)}
  {\Pr(s^0 \xleftarrow{x^1} s^1 \cdots s^{N-2} \xleftarrow{x^{N-1}} s^{N-1} | s^{N-1} , \xR )}
  & = e^{-\beta Q_\text{F}}
  ~.
\end{align*} 
Otherwise, microscopic reversibility is broken.

Although, microscopic reversibility has also been referred to as a ``detailed
fluctuation theorem'', it is actually an assumption appropriate only in special
cases. For example, Eq.~\eqref{eq:embqAsProduct} shows that if the dynamics are
Markovian over states (given input) and obey detailed balance (\`{a} la \S
\ref{sec:DetailedBalance}), then microscopic reversibility is satisfied for
arbitrary non-Markovian inputs. In essence, this is the justification of
microscopic reversibility suggested by Crooks \cite{Crooks98, Crooks99} from
which his eponymous fluctuation theorem follows.

In this view, detailed balance and microscopic reversibility are effectively
the same assumption since each implies the other. Section \S~\ref{sec:GenCFT}
generalizes the CFT to describe fluctuation laws in the absence of microscopic
reversibility.

\subsection{Path dependence and irreversibility}
\label{sec:Irreversibility}

The importance of state-space path dependence is captured via an informational
quantity $\Upsilon$ we call the \emph{path relevance} of a state sequence
$s_{1:N}$ given initial state $s_0$ and input sequence $x_{1:N}$: 
\begin{align}
\Upsilon(s_{1:N} | s_0 , x_{1:N} )
  \equiv \ln
  \frac{ \Pr(s_{1:N} | s_0 , x_{1:N} ) }
  { \prod_{n=1}^{N-1} \pi_{x_n}(s_{n}) }
  ~.
\label{eq:PathRelevanceDef}
\end{align}
(The branching Pythagorean letter $\Upsilon$ recognizes its ancient
symbolism---divergent consequences of choosing one path over another.) Note that the equilibrium
probabilities in the denominator do not depend on the original state, whereas
the numerator (even after factoring) depends on state-to-state transitions. As
the environmental input drives the system probability density through its
state-space, path relevance develops in the state sequence. A joint sequence
lacks path relevance, if $\Upsilon=0$ for that sequence.

Whenever state transitions are Markovian given the input, the numerator in Eq.~\eqref{eq:PathRelevanceDef} simplifies to:
\begin{align*}
\Pr(s_{1:N} | x_{1:N}, s_0) = \prod_{n=1}^{N-1} \Pr(s_n | s_{n-1}, x_n)
  ~,
\end{align*}
and the path relevance becomes: 
\begin{align*}
\Upsilon = \sum_{n=1}^{N-1}
  \ln \frac{\Pr(s_n | s_{n-1}, x_n)}{\pi_{x_n}(s_{n})}
  ~.
\end{align*}
Thus, there is path relevance even for Markov processes. The actual driving
history matters. When a system is \emph{non}-Markovian, there are yet
additional contributions to path relevance.

Path relevance of a particular state sequence given a particular driving is a
system feature, regardless of the environment in which the system finds itself.
However, expectation values involving the above relationship can reflect the
environment's nature.

For our development, we find it useful to consider both the forward-path
dependence and the reverse-path dependence of a particular joint sequence: $x^1 \dots x^{N-1}$ and $s^0 \dots s^{N-1}$.
The \emph{forward-path dependence} is as expected:
\begin{align}
\Upsilon & = \Upsilon (\sF | s^0 , \xF) \nonumber \\
  & = \ln \frac{ \Pr(\sF | s^0 , \xF) } { \prod_{n=1}^{N-1} \pi_{x^n}(s^{n}) }
  ~,
\end{align}
and, similarly, the \emph{reverse-path dependence} is:
\begin{align}
\RPD & = \RPD (\sF | s^{0} , \xF ) \nonumber \\
  & = \Upsilon (\sR | s^{N-1} , \xR ) \nonumber \\
  & = \ln \frac{ \Pr(\sR | s^{N-1} , \xR ) } { \prod_{n=1}^{N-1} \pi_{x^{n}}(s^{n-1}) } ~.
\end{align}

And, finally, we have the \emph{irreversibility}:
\begin{align}
\Psi \equiv \Upsilon - \RPD
  ~,
\end{align}
the \emph{net directional relevance}---of a particular path $\sF$ given
$s^0$ and $\xF$.
Nonzero $\Psi$ indicates the irrevocable consequences of path traversal.
\emph{Microscopically reversible dynamics have $\Psi = 0$ for all paths with
nonzero probability}, indicating no divergence in path branching anywhere
through the state-space. And so, $\Psi = 0$ for all paths with nonzero
probability if and only if the dynamic satisfies detailed balance. In short,
$\Psi$ quantifies the imbalance in path reciprocity along a driven
state-sequence.

Sometimes $\RPD$ can be $-\infty$ for an allowed forward path $\Pr(\sF | s^0 ,
\xF) > 0$, corresponding to a forbidden reverse path $\Pr(\sR | s^{N-1} , \xR)
= 0$. This is a situation that never arises with detailed balance dynamics.
Such paths are infinitely irreversible: $\Psi = \infty$.

\section{Generalized fluctuation theorems for nonequilibrium systems}
\label{sec:FTs}

Absent microscopic reversibility, the architecture of transitions over
state-space matters. More concretely, we will constructively show how this
architecture affects the nonequilibrium thermodynamics of complex systems.

\subsection{Generalized Detailed Fluctuation Theorem}
\label{sec:GenDetailedFT}

Assume the system under study starts from some distribution
$\boldsymbol{\mu}_\text{F}$ and that the associated reverse trajectory (when
starting from some other distribution $\boldsymbol{\mu}_\text{R}$) is
allowed---that is, it has nonzero probability. Then the ratio of conditional
probabilities of a state sequence (given a driving sequence) to the reversed
state sequence (given reversed driving) is:
\begin{widetext}
\begin{align}
&
\frac{\Pr(\StartF \xrightarrow{x^0} s^0 \xrightarrow{x^1} s^1 \cdots s^{N-2}
\xrightarrow{x^{N-1}} s^{N-1}   | \StartF , x^0 \xF )}
  {\Pr(s^0 \xleftarrow{x^1}
s^1 \cdots s^{N-2} \xleftarrow{x^{N-1}} s^{N-1} \xleftarrow{x^N} \StartR  |
\StartR , x^N \xR )} \nonumber \\
  & \qquad \qquad \qquad \qquad \qquad \qquad \quad =
  \frac{\Pr(\St_{0:N} = s^0 \sF  | \St_{-1} \sim \StartF , X_{0:N} = x^0 \xF )}{\Pr(\St_{0:N} = s^{N-1} \sR | \St_{-1} \sim \StartR , X_{0:N} = x^N \xR )} 
 \nonumber \\
& \qquad \qquad \qquad \qquad \qquad \qquad \quad =  \frac{\Pr(s^0 | \StartF , x^0 )}{\Pr(s^{N-1} | \StartR , x^N )} \, \,
    \frac{\Pr( \sF | s^0 , \xF )}{\Pr( \sR | s^{N-1} , \xR )} \nonumber \\
& \qquad \qquad \qquad \qquad \qquad \qquad \quad = 
    \frac{\Pr(s^0 | \StartF , x^0 )}{\pi_{x^0}(s^0)} \, \,
    \frac{\pi_{x^{N}}(s^{N-1})}{\Pr(s^{N-1} | \StartR , x^{N} )} \, \,
    \frac{\Pr( \sF | s^0 , \xF )}{\prod_{n=1}^{N-1} \pi_{x^n}(s^n) } \, \,
    \frac{ \prod_{n=1}^{N-1} \pi_{x^n}(s^{n-1}) }{\Pr( \sR | s^{N-1} , \xR )} \, \, 
    \prod_{n=0}^{N-1} \frac{ \pi_{x^n}(s^n) }{ \pi_{x^{n+1}}(s^n) } \nonumber \\
& \qquad \qquad \qquad \qquad \qquad \qquad \quad = 
    \frac{\Pr(s^0 | \StartF , x^0 )}{\pi_{x^0}(s^0)} \, \,
    \frac{\pi_{x^{N}}(s^{N-1})}{\Pr(s^{N-1} | \StartR , x^{N} )} \, \,
    e^{\Psi_\text{F}} \, e^{\eeep_\text{F}} \nonumber \\
& \qquad \qquad \qquad \qquad \qquad \qquad \quad = 
    e^{\gamma(s^0 | \StartF , x^0 ) - \gamma(s^{N-1} | \StartR , x^{N} )} \, \,
    e^{\eeep_\text{F} + \Psi_\text{F}} ~,
    \label{eq:GenDetailedFT}
\end{align}
\end{widetext}
where $\eeep_\text{F} = \eeep(X_{0:N+1} = x^0 \xF x^N, \St_{0:N} = s^0 \sF)$ is
the excess environmental entropy production in the forward trajectory,
$\Psi_\text{F} = \Psi(\St_{0:N} = s^0 \sF | X_{1:N} = \xF)$ is the
irreversibility of the forward trajectory, and $\gamma(s | \boldsymbol{\mu} ,
x) = \ln \left(\Pr(s | \boldsymbol{\mu} , x ) / \pi_{x}(s) \right)$ is the
nonsteady-state addition to free energy associated with being in the
nonsteady-state distribution $\boldsymbol{\mu}$ with environmental drive $x$.

Since $\Psi_\text{F}$ can diverge for forward paths with nonzero probability, 
we typically rewrite Eq.~\eqref{eq:GenDetailedFT} as the ``less divergent'' expression:
\begin{align}
e^{- \gamma_\text{F}} \Pr(s^0 \sF  & |  \StartF , x^0 \xF ) e^{- \Psi_\text{F}}
  \nonumber \\
  & = e^{- \gamma_\text{R}}
  \Pr( s^{N-1} \sR |  \StartR ,  x^N \xR ) e^{ \eeep_\text{F}}
  ~.
  \label{eq:FundamentalDFT}
\end{align}
Eq.~\eqref{eq:FundamentalDFT} is the fundamental relation for all that follows:
it relates the probabilities of forward and reverse trajectories via entropy
production $\eeep_\text{F}$ of the forward path, irreversibility
$\Psi_\text{F}$ of the forward path, and change $\beta^{-1} ( \gamma_\text{F} -
\gamma_\text{R} )$ in the nonsteady-state addition to free energy between the
forward and reverse start-distributions.

In what follows it will be all too easy to write seemingly divergent
expressions. Such divergences do not manifest themselves when taking
expectation values for physical quantities involving them, since they come
weighted with zero probability. This is similar to the reasonable convention
for Shannon entropies that $0 \log 0 = 0$. Nevertheless, caution is advised
when probabilities vanish.

\subsection{Simplifications}
\label{sec:Simplify}

Before proceeding and to aid understanding, let's consider several special
cases. If the forward drive or protocol begins with the system equilibrated
to the static environmental drive $x^0$, 
then $\StartF = \boldsymbol{\pi_{x^0}}$ and $\gamma(s^0 | \StartF , x^0 ) = 0$.
Similarly, if the reverse protocol begins with the system equilibrated to the
static environmental drive $x^N$, 
then $\StartR = \boldsymbol{\pi_{x^N}}$ and $\gamma(s^N | \StartR , x^N ) = 0$.
In this case, Eq. (\ref{eq:FundamentalDFT}) simplifies to:
\begin{align*}
\frac{\Pr(\boldsymbol{\pi_{x^0}} \xrightarrow{x^0} s^0 \cdots \xrightarrow{x^{N-1}} s^{N-1} | \boldsymbol{\pi_{x^0}} , x^0 \xF )}
  {\Pr(s^0 \xleftarrow{x^1} \cdots s^{N-1} \xleftarrow{x^N} \boldsymbol{\pi_{x^N}}  | \boldsymbol{\pi_{x^N}} , x^N \xR )}
=
    e^{\eeep_\text{F} + \Psi_\text{F}} ~.
\end{align*}

As a separate matter, if the dynamics are microscopically reversible, then
$\Psi = 0$. Consider the very special case where (i) the dynamics are
microscopically reversible, (ii) the forward driving begins with the system
equilibrated with $x^0$, and (iii) the reverse driving begins with the system
equilibrated with $x^N$. Then, the ratio of probabilities of observing a
forward state sequence (given forward driving) and observing the reversal of
that state sequence (given the reversal of that driving) is simply
$e^{\eeep_\text{F}}$. That is, the forward sequence is exponentially more
likely if it has positive entropy production.

Apparently, the more general case is more nuanced and, beyond depending on a
nonsteady-state starting distribution, it depends strongly on the architecture
of branching transitions among states.

Another interesting special case is if $x^{N} = x^{N-1}$ and $\StartR$ is the
distribution that the forward driving induces from $\StartF$. Then the
\emph{dissipated work} $W_\text{diss} \equiv \Wex - \beta^{-1} \Delta \gamma$
associated with the forward trajectory comes into play. (Recall that
$\beta^{-1} \Delta \gamma$ is the change in nonsteady-state contributions to
free energy.) Then the ratio of forward- and reverse-path probabilities is:
\begin{align*}
& \frac{\Pr(\StartF \xrightarrow{x^0} \cdots \xrightarrow{x^{N-1}} s^{N-1}
  | \StartF , \xF )}
  {\Pr(s^0 \xleftarrow{x^1} \cdots \xleftarrow{x^N} \boldsymbol{\mu}(\StartF , \xF)
  | \boldsymbol{\mu}(\StartF , \xF) , \xR )} \\
  & \qquad \qquad \qquad \qquad \qquad \qquad \qquad = e^{\Psi_\text{F}} e^{\beta [\Wex - \beta^{-1} \Delta \gamma] } \\
  & \qquad \qquad \qquad \qquad \qquad \qquad \qquad = e^{\Psi_\text{F}} e^{\beta W_\text{diss} } 
  ~.
\end{align*}
Even in the case of microscopic reversibility, this is useful since it
generalizes previous FTs to nonequilibrium start and end distributions.  In the
case of microscopic reversibility, $\Psi_\text{F} = 0$ and so the ratio of
forward- and reverse-path probabilities from any nonequilibrium start and end
distribution is exponential $e^{\beta W_\text{diss}}$ in the dissipated work.
Thus, an experimental test of this result is one with time-symmetric driving.
The forward protocol corresponds to the first half of the driving while the
reverse protocol is the second half. Clearly, the final nonequilibrium
distribution for the forward protocol is the same as the initial nonequilibrium
distribution for the reverse protocol. The dissipated work then corresponds to
that dissipated in the first half of the driving. Practically, in cases where
the dynamic is not microscopically reversible, this allows experimentally
extracting the system's irreversibility $\Psi$.

\subsection{Generalized Crooks Fluctuation Theorem}
\label{sec:GenCFT}

We can now turn to the irreversible analog of the Crooks Fluctuation Theorem
(CFT).

First, we note that both entropy production $\eeep$ and irreversibility
$\Psi$ are odd under time reversal. Explicitly, we have:
\begin{align*}
\eeep (X_{0:N+1} & = x^0 \xF x^N, S_{0:N} = s^0 \sF) \\
    & = \ln  \prod_{n = 0}^{N-1}
	\frac{\pi_{x^n}(s^{n})}{\pi_{x^{n+1}}(s^{n})}  \\
	& =  - \ln  \prod_{n = 0}^{N-1}
	\frac{\pi_{x^{n+1}}(s^{n})}{\pi_{x^n}(s^{n})}   \\
	& = - \ln  \prod_{n = 0}^{N-1}
	\frac{\pi_{x^{N-n}}(s^{N-1-n})}{\pi_{x^{N-1-n}}(s^{N-1-n})}  \\
    & = - \eeep(X_{0:N+1} = x^N \xR x^0, S_{0:N} = s^{N-1} \sR)
\end{align*}
and:
\begin{align*}
\Psi (S_{0:N} & = s^0 \sF | X_{1:N} = \xF) \\
	& = \ln \Bigl[ \frac{\Pr(\sF | s^0 , \xF)}{\Pr(\sR | s^{N-1} , \xR)} \prod_{n=1}^{N-1} \frac{\pi_{x^n}(s^{n-1})}{\pi_{x^n}(s^{n})} \Bigr] \\
	& = - \ln \Bigl[ \frac{\Pr(\sR | s^{N-1} , \xR)}{\Pr(\sF | s^0 , \xF)}  \prod_{n=1}^{N-1} \frac{\pi_{x^n}(s^{n})}{\pi_{x^n}(s^{n-1})} \Bigr] \\
	&= - \Psi(S_{0:N} = s^{N-1} \sR | X_{1:N} = \xR)
	~.
\end{align*}
For brevity, let 
$\eeep_\text{F} \equiv \eeep(X_{0:N+1} = x^0 \xF x^N, \St_{0:N} = s^0 \sF)$ and
$\eeep_\text{R} \equiv \eeep(X_{0:N+1} = x^N \xR x^0, \St_{0:N} = s^{N-1}
\sR)$.
And, similarly, 
$\Psi_\text{F} \equiv \Psi(\St_{0:N} = s^0 \sF | X_{1:N} = \xF)$ and
$\Psi_\text{R} \equiv \Psi(\St_{0:N} = s^{N-1} \sR | X_{1:N} = \xR)$. 
In this notation, we just established that $\eeep_\text{F} = - \eeep_\text{R}$
and $\Psi_\text{F} = -\Psi_\text{R}$.

Second, if we now choose $\StartF = \boldsymbol{\pi_{x^0}}$ and $\StartR = \boldsymbol{\pi_{x^N}}$ and marginalize over all possible state trajectories, we find that the joint probability of entropy production and irreversibility given the driving protocol starting from an equilibrium distribution is:
\begin{align*}
\Pr( & \eeep ,  \Psi | \boldsymbol{\pi_{x^0}}, x^0 \xF x^N) \\
&= \sum_{s^0 \sF \in \SSet^N} 
  \Pr(s^0 \sF  | \boldsymbol{\pi_{x^0}} ,  x^0 \xF ) \, \delta_{\eeep , \eeep_\text{F}} \delta_{\Psi , \Psi_\text{F} } \\
&= \sum_{s^0 \sF \in \SSet^N}  e^{\eeep_\text{F}} e^{\Psi_\text{F}} 
  \Pr( s^{N-1} \sR | \boldsymbol{\pi_{x^N}} ,  x^N \xR ) \, \delta_{\eeep , \eeep_\text{F}} \delta_{\Psi , \Psi_\text{F} } \\
&= e^{\eeep} e^{\Psi} \sum_{s^0 \sF \in \SSet^N}  
  \Pr( s^{N-1} \sR | \boldsymbol{\pi_{x^N}} ,  x^N \xR ) \, \delta_{\eeep , \eeep_\text{F}} \delta_{\Psi , \Psi_\text{F} } \\
&= e^{\eeep} e^{\Psi} \!\!\!\! \sum_{s^{N-1} \sR \in \SSet^N}  \!\!\!\!
  \Pr( s^{N-1} \sR | \boldsymbol{\pi_{x^N}} ,  x^N \xR ) \, \delta_{\eeep , - \eeep_\text{R}} \delta_{\Psi , - \Psi_\text{R} } \\
&= e^{\eeep} e^{\Psi} \Pr(- \eeep , - \Psi | \boldsymbol{\pi_{x^N}}, x^N \xR x^0) ~.
\end{align*}

Finally, we rewrite this to give the \emph{extended CFT for irreversible
processes}:
\begin{align}
\frac{\Pr(\eeep , \Psi | \boldsymbol{\pi_{x^0}}, x^0 \xF x^N)}{\Pr(- \eeep , - \Psi | \boldsymbol{\pi_{x^N}}, x^N \xR x^0)} = e^\Psi e^\eeep
  ~.
\label{eq:genCFT}
\end{align}

\subsection{Interpretation}
\label{sec:InterpCFT}

In the special case of isothermal time-symmetric driving---$x^0 \xF x^0 = x^0
\xR x^0 = x^0 x^1 x^2 \dots x^2  x^1 x^0$---and starting from an equilibrium
distribution, Eq.~\eqref{eq:genCFT} provides a useful comparison between values
of excess work achieved by the single time-symmetric driving protocol:
\begin{align}
\frac{\Pr(\Wex , \Psi )}{\Pr(- \Wex , - \Psi )} = e^\Psi e^{\beta \Wex }
  ~.
\label{eq:TSgenCFTwW}
\end{align}

Equation~\eqref{eq:genCFT} should be compared to the original CFT that, in its
most general form, can be written (with necessary interpretation) as
\cite{Crooks00}:
\begin{align}
\frac{\Pr_\text{F}(\eeep)}{\Pr_\text{R}(-\eeep)} = e^\eeep ~.
\label{eq:StandardCFT}
\end{align}
It is tempting to write Eq.~\eqref{eq:StandardCFT} as:
\begin{align}
\frac{\Pr(\eeep | \boldsymbol{\pi_{x^0}}, x^0 \xF x^N)}{\Pr(-\eeep |
\boldsymbol{\pi_{x^N}}, x^N \xR x^0)} \, \overset{?}{=} \, e^\eeep
  ~.
\label{eq:StandardCFTInterpreted}
\end{align}
This form presents some concerns, however. In the case of detailed balance, though,
$\Psi = 0$ for all trajectories, and so our Eq.~\eqref{eq:genCFT} guarantees
Eq.~\eqref{eq:StandardCFTInterpreted} in the case of detailed balance. Crooks'
original CFT derivation \cite{Crooks98,Crooks99} also assumed detailed balance,
and so Eq.~\eqref{eq:StandardCFTInterpreted} was implied.

However, absent detailed balance, Eq.~\eqref{eq:StandardCFT} has a rather
different interpretation: $\Pr_\text{R}(\cdot)$ then implies not only the
reversed driving, but also that the distribution describes a different
``reversed'' system that is not of direct physical
relevance~\cite{Crooks00,Cher06}. One consequence is that the probabilities in
the numerator and denominator are not comparable in any physical sense. So, in
general, we have:
\begin{align}
\frac{\Pr(\eeep | \boldsymbol{\pi_{x^0}}, x^0 \xF x^N)}{\Pr(-\eeep | \boldsymbol{\pi_{x^N}}, x^N \xR x^0)} \neq e^\eeep ~.
\label{eq:CFTviolation}
\end{align}
In contrast, our irreversible CFT in Eq.~\eqref{eq:genCFT} compares
probabilities of entropy production (and path irreversibility) for the same
thermodynamic system under a control protocol and under the reversed control
protocol. Equation~\eqref{eq:genCFT}, unlike equalities involving an unphysical
dual dynamic as in Eq. (\ref{eq:StandardCFT}), allows a clear and meaningful
physical interpretation of the relationship between entropies produced and,
moreover, is not limited by assuming detailed balance.

Note that our Eq.~\eqref{eq:genCFT}, expressed in terms of excess environmental
entropy production $\eeep$ and path irreversibility $\Psi$, does not make
explicit mention of temperature. Indeed, if temperature dependence is folded
into different environmental inputs $x$, then Eq.~\eqref{eq:genCFT} applies
just as well to systems driven by environments with spatially inhomogeneous
temperature distributions that change in time. Explicitly, $\boldsymbol{\pi_x}$
and $\boldsymbol{\pi_{x'}}$ could represent the distribution over effective
states induced by environmental conditions associated with $x$ and $x'$
\emph{including} their different spatial distributions of temperature.

\subsection{Translation to Steady-State Thermodynamics}
\label{sec:SST_FTs}

A better understanding of the irreversible CFT comes by comparing it to recent
related work. Most directly, our results complement those on driven transitions
between NESSs. Specifically, the importance of nondetailed-balanced dynamics in
enabling the organization of complex nonequilibrium behavior has been
considered previously. For example, Ref. \cite{Gaveau97} also introduced a
path entropy which is an ensemble average of that considered here.

Another comparison is found in Ref. \cite{Hatano01}'s nonequilibrium
thermodynamics over NESSs using housekeeping $Q_\text{hk}$ and excess
$Q_\text{ex}$ heats. While that treatment focused on Langevin dynamics,
we find that in general $Q_\text{hk}$ corresponds directly to our path
irreversibility $\Psi$. Specifically, in the isothermal setting there,
according to Eq.~(35), we have:
\begin{align*}
\beta Q_\text{hk} \approx \Psi 
  ~.
\end{align*} 
Indeed, for isothermal Markovian dynamics Eq.~(7.7) of Ref.~\cite{Harris07}
suggests (via their Eqs. (2.11) and (7.1)) that this is in fact an equality:
\begin{align}
\beta Q_\text{hk} = \Psi
  ~.
\end{align} 
Reference~\cite{Espo10} called the irreversibility $\Psi$ the \emph{adiabatic
contribution} to entropy production. Several related translations from Ref.
~\cite{Hatano01} to our setting can also be easily made:
$\rho_\text{ss}(s;x) \to \pi_x(s)$,
$\phi(s; x) \to -\log \pi_x(s)$,
$\Delta S \to \Delta S_\text{ss} $, 
and 
$\beta Q_\text{ex} + \Delta \phi \to \eeep$.
Hence, $\braket{\eeep} \geq 0$ (for Langevin systems) is Ref.
~\cite{Hatano01}'s main result. From these connections, we see that our
development not only provides new constraints on detailed fluctuations, but
also extends these earlier results beyond Langevin systems.

Exposing these translations allows reformulating our detailed fluctuation
theorems to steady-state thermodynamics (SST). We have:
\begin{align*}
e^{\gamma(s^0 | \StartF , x^0 ) - \gamma(s^{N-1} | \StartR , x^{N} )} \, \,
  e^{\eeep_\text{F} + \Psi_\text{F}} 
  & = e^{\beta(Q_\text{ex} + Q_\text{hk}) + \Delta S^\text{sys} } \\
  & = e^{\Delta S^\text{tot}_\text{F}}
  ~,    
\end{align*}
where $\Delta S^\text{sys} \equiv -\ln \frac{\Pr( s^{N-1} | \StartR , x^{N} )}{ \Pr( s^{0} | \StartF , x^{0} ) }$
when we choose $\bra{\StartR} = \bra{\StartF} \T^{(\SSet \to \SSet| x^0 \xF)}$
and where $S^\text{tot}_\text{F}$ is the total change in entropy in forward
time. This yields:
\begin{align*}
\frac{\Pr(\St_{0:N} = s^0 \sF  | \St_{-1} \sim \StartF , X_{0:N} = x^0 \xF )}{\Pr(\St_{0:N} = s^{N-1} \sR | \St_{-1} \sim \StartR , X_{0:N} = x^N \xR )} 
&= e^{\Delta S^\text{tot}_\text{F}}
  ~.
\end{align*}
And so, we immediately see that:
\begin{align*}
\braket{ e^{- \Delta S^\text{tot}_\text{F}} }_{\Pr(\St_{0:N} = s^0 \sF  | \St_{-1} \sim \StartF , X_{0:N} = x^0 \xF )}
  = 1
  ~.
\end{align*}
This extends the validity of Ref. \cite{Seifert05}'s general integral
fluctuation theorem beyond Langevin dynamics. Since the total change in entropy
is time asymmetric---$\Delta S^\text{tot}_\text{F} = - \Delta
S^\text{tot}_\text{R}$---we obtain the most direct CFT generalization valid
outside of detailed balance:
\begin{align}
\frac{\Pr( \Delta S^\text{tot} | \boldsymbol{\pi_{x^0}}, x^0 \xF x^N)}{\Pr(- \Delta S^\text{tot} | \boldsymbol{\pi_{x^N}}, x^N \xR x^0)} = e^{\Delta S^\text{tot}} ~.
\label{eq:TotS_DFT}
\end{align}
Again, this does not invoke a dual, unphysical dynamic.
Equation~\eqref{eq:TotS_DFT} has been reported previously in various settings;
see, e.g., Eq.~(21) of Ref.~\cite{Espo10} and Eq.~(43) of Ref.~\cite{Cher06}.
The result gives a detailed fluctuation relation for the change in total
entropy production when transitioning between steady states.

The new detailed fluctuation theorem of Eq. (\ref{eq:genCFT}) for joint
distributions goes further in refining SST. If starting in a steady state and
executing a protocol in an isothermal environment, we find that:
\begin{align*}
\frac{\Pr( \Wex , Q_\text{hk} | \boldsymbol{\pi_{x^0}}, x^0 \xF x^N)}{\Pr(- \Wex , - Q_\text{hk} | \boldsymbol{\pi_{x^N}}, x^N \xR x^0)}
  = e^{\beta Q_\text{hk} } e^{\beta \Wex}
  ~.
\end{align*}
This novel relation gives strong constraints on the thermodynamic behavior of
systems driven between NESSs, since it constrains the \emph{joint} distribution
for excess work and housekeeping heat. Moreover, nonsteady-state additions
to free energy are predicted when an experiment does not start in steady state.

In the special case of time-symmetric driving---$x^0 \xF x^0 = x^0 \xR x^0 =
x^0 x^1 x^2 \dots x^2  x^1 x^0$---and starting from an equilibrium
distribution, the preceding expression reduces to a useful comparison between
excess work values achieved by the single time-symmetric protocol: 
\begin{align*}
\frac{\Pr(\Wex , \Qhk  )}{\Pr(- \Wex , - \Qhk )}
  = e^{\beta \Qhk } e^{\beta \Wex}
  ~.
\end{align*}
Similar results were recently derived in Ref. \cite{Lahiri14} under more
restrictive assumptions for underdamped Langevin systems.

This all said, one must use caution and not always identify $\Psi$ with $\beta
\Qhk$. Most importantly, not all sources of irreversibility are naturally
characterized as ``heat''. Thinking of irreversibility on its own dynamical
terms is best.

\subsection{Integral fluctuation theorems}
\label{sec:IntegralFTs}

Integral fluctuation theorems in the absence of detailed balance, starting
arbitrarily far from equilibrium, also follow straightforwardly. One
generalization of the integral fluctuation theorem \cite{Jarz97a} is:
\begin{align}
& \left\langle
  e^{-\beta W_\text{diss} - \Psi}  \right\rangle_{\Pr( s_{0:N} | \StartF, \xF) }
  \nonumber \\
  & \qquad = \sum_{s_{0:N} \in \SSet^N} \Pr( s_{0:N} | \StartF, \xF)
  e^{-\beta W_\text{diss} - \Psi} \nonumber \\
  & \qquad = \sum_{s_{0:N} \in \SSet^N} \Pr(s^0 \xleftarrow{x^1} \cdots
  \xleftarrow{x^N} \boldsymbol{\mu}(\StartF , \xF)
  | \boldsymbol{\mu}(\StartF , \xF) , \xR ) \nonumber \\
  & \qquad = 1
  ~.
\label{eq:GenIFT}
\end{align}
If the input is stochastic, then averaging over the input also gives:
\begin{align*}
\left\langle  e^{-\beta W_\text{diss} - \Psi} \right\rangle_{\Pr( x_{0:N} , s_{0:N} | \StartF) } = 1
  ~.
\end{align*}
Note that this relation does \emph{not} require the system to be in steady
state at any time.

From the concavity of the exponential function, it is tempting to assert a corresponding generalized Second Law of SST as:
\begin{align}
\left\langle W_\text{diss} \right\rangle \geq - \left\langle  \Qhk  \right\rangle  ~.
\label{eq:WdissvsIRR}
\end{align}
Although Eq.~\eqref{eq:WdissvsIRR} is true, notably it is neither a strong nor
useful bound. Let's address this. Note that: 
\begin{align}
\left\langle  e^{- \Psi}  \right\rangle_{\Pr( s_{0:N} | \StartF, \xF) } & = 1
\label{eq:PsiIFT}
\end{align}
and:
\begin{align}
\left\langle  e^{-\beta W_\text{diss}}
  \right\rangle_{\Pr( s_{0:N} | \StartF, \xF) }
  & = 1
  ~.
\label{eq:WdissIFT}
\end{align}
Both follow from the normalization of probabilities of the conjugate dynamic.
Therefore, $\braket{\Psi} \geq 0$ \emph{and} $\braket{W_\text{diss}} \geq 0$.
And, hence $\braket{\Qhk} \geq 0$ as shown in Ref.~\cite{Speck05}. So,
Eq.~\eqref{eq:WdissvsIRR} is devoid of utility. Nevertheless,
Eq.~\eqref{eq:GenIFT} puts a novel constraint on the joint distributions of
$W_\text{diss}$ and $\Psi$.

Introducing an artificial conjugate dynamic following Ref.~\cite{Crooks00} and
following the derivation there with $\phi / \beta$ in place of $E$, when
starting in the steady state distribution $\boldsymbol{\pi_{x_0}}$, we can show
that:
\begin{align}
\left\langle  e^{-\eeep }  \right\rangle_{\Pr( s_{0:N} | \boldsymbol{\pi_{x_0}}, \xF) }  = 1 ~,
\label{eq:BasicIFT}
\end{align}
which implies the restriction $\braket{\eeep} \geq 0$. Despite similar
appearance, this result has meaning beyond Crooks' derivation of the Jarzynski
equality, as it now also applies \emph{atop nonequilibrium steady states}.
Recall that $\eeep$ has general meaning as in Eq.~\eqref{eq:EEEPdef}: $\eeep =
\beta \Wex = -\beta \Qex + \Delta \phi$. So, Eq. \eqref{eq:BasicIFT} becomes:
\begin{align}
\left\langle  e^{-\beta \Wex }  \right\rangle_{\Pr( s_{0:N} | \boldsymbol{\pi_{x_0}}, \xF) }  = 1 ~.
\end{align}
Effectively, this is Ref. \cite{Hatano01}'s relation that, with our sign convention for $\Qex$, implies:
\begin{align}
\braket{\eeep} & = \braket{-\beta \Qex + \Delta \phi} \nonumber \\
               & \geq 0 
\label{eq:HSinequality}
  ~,
\end{align}
for processes that start in steady state.

However, using Eq.~\eqref{eq:WdissIFT}, we find a more precise constraint on expected excess entropy production whether or not the system starts in steady state:
\begin{align}
\braket{\eeep} \geq \Delta \braket{ \gamma } ~,
\end{align}
where the RHS can be positive or negative, but can only be negative if the
system starts out of steady state. When starting in a steady state, this yields:
\begin{align}
\braket{\eeep} 
  & \geq \braket{ \gamma_\text{final}} \\
  & = D_\text{KL}
  \bigl[ \Pr(\St_t  | \St_{0} \sim \boldsymbol{\pi_{x_0}} ,  x_{1:t+1} ) \, || \, \boldsymbol{\pi_{x_t}} \bigr]
  ~,
\end{align}
which
is a stronger constraint than the previous result of
Eq.~\eqref{eq:HSinequality}, since the RHS is always positive for $ \Pr(\St_t
| \St_{0} \sim \boldsymbol{\pi_{x_0}} , x_{1:t+1} )
\neq\boldsymbol{\pi_{x_t}}$.

Integral fluctuation theorems for systems with controlled or intrinsic feedback
also directly follow, as we now show, extending the theory of feedback control
to the setting of transitions between NESSs.

\subsection{Fluctuation theorems with an auxiliary variable}
\label{sec:FeedbackFTs}

Actions made by a complex thermodynamic system can couple back from the
environment to influence the system's future input. To achieve this, the system may
express an auxiliary random variable $Y_t$---the current ``output'' that takes
on the values $y \in \mathcal{Y}$ and is instantaneously energetically mute,
but may influence the future input and so does have energetic relevance.

The variable $Y_t$ could be measurement, output, or any other auxiliary
variable that influences the state or input sequences. To be concise, we
introduce a shorthand for the time-ordered sequences of random variables:
$\vXF \equiv X_{0:N}$, $\vSF \equiv \St_{0:N}$, and $\vYF \equiv
Y_{0:N}$. And, for particular realizations of the sequences: $\vxF \equiv
x^{0:N}$, $\vsF \equiv s^{0:N}$, and $\vyF \equiv y^{0:N}$. When time reversing
realizations, we let $\vxR = x^{N-1} x^{N-2} \cdots x^1 x^0$ and $\vsR =
s^{N-1} s^{N-2} \cdots s^1 s^0$. To clarify further, $\vxF$ appearing inside a
probability implies $\vXF = \vxF$ and $\vsR$ appearing inside a probability
implies $\vSR = \vsR$.

We quantify how much the auxiliary variable is independently informed from the
state sequence---beyond what could be known if given only the initial
distribution over states and the driving history---via the unaveraged
conditional mutual information:
\begin{align*}
i(\vsF; \vyF | \vxF , \StartF )
  & \equiv \ln
  \frac{\Pr(\vsF , \vyF | \vxF , \StartF )}{ \Pr( \vyF | \vxF , \StartF )
  \Pr( \vsF | \vxF , \StartF ) }
  \\
  & = \ln 
  \frac{\Pr(\vsF , \vyF , \vxF | \StartF )}{ \Pr( \vyF , \vxF | \StartF )
  \Pr( \vsF | \vxF , \StartF ) }
  ~.
\end{align*}
Note that averaging over the input, state, and auxiliary sequences gives the familiar conditional mutual information:
$\I[\vSF ; \vYF | \vXF , \StartF ] = \left\langle i[\vsF ; \vyF | \vxF ,
\StartF ] \right\rangle_{\Pr(\vxF , \vsF , \vyF | \StartF )}$.  

As detailed in the App.~\ref{sec:FTappendix2}:
\begin{align*}
e^{\beta W_\text{diss}
  + i[\vsF ; \vyF | \vxF , \StartF ] + \Psi } 
  & = \frac{\Pr(\vsF , \vyF , \vxF | \StartF )}
  { \Pr( \vyF , \vxF | \StartF ) \Pr( \vsR | \vxR , \StartR ) }
  ~,
\end{align*}
where $\StartR = \boldsymbol{\mu}(\StartF, \vxF)$.
This leads directly to the integral fluctuation theorem:
\begin{align}
\left\langle  e^{-\beta  W_\text{diss} - i[\vsF; \vyF | \vxF , \StartF ] - \Psi
} \right\rangle_{\Pr(\vsF , \vyF , \vxF | \StartF )} 
 = 1
  ~.
\end{align}
However, as before, the resulting bound on $\braket{W_\text{diss}}$ is not the
tightest possible. Alternatively, we can invoke the normalization of conjugate
dynamics to show:
\begin{align}
\left\langle  e^{-\beta  W_\text{diss} - i[\vsF; \vyF | \vxF , \StartF ] }
\right\rangle_{\Pr(\vsF , \vyF , \vxF | \StartF )} 
& = 1 ~.
\end{align}
This implies a new lower bound for the revised Second Law of Thermodynamics:
\begin{align}
\left\langle W_\text{diss} \right\rangle  
\ge -  \kB T \, \I[\vSF; \vYF | \vXF , \StartF ]   ~,
\end{align}
enabled by the conditional mutual information between state-sequence and
auxiliary sequence, given input-sequence. Notably, this relation holds
arbitrarily far from equilibrium and allows for the starting and ending
distributions to be nonsteady-state.

We may also be interested in the unaveraged unconditioned mutual information
between the auxiliary variable sequence and the joint input--state sequence.
Then, using:
\begin{align*}
i[\vyF ; \vxsF |  \StartF ]
  \equiv \ln 
  \frac{\Pr(\vxF , \vsF , \vyF |  \StartF )}
  { \Pr( \vyF |  \StartF ) \Pr( \vxF , \vsF |  \StartF ) }
  ~,
\end{align*}
we find that, in general:
\begin{align}
\left\langle W_\text{diss} \right\rangle  \ge -  \kB T \,
  \I[\vYF; \vXSF |  \StartF ]
\end{align}
and when starting in steady-state:
\begin{align}
 \left\langle \eeep \right\rangle  
\ge - \I[\vYF; \vXSF |   \boldsymbol{\pi}_{x^0} ]
~.
\end{align}

One can now continue in this fashion to successively derive a seeming unending
sequence of fluctuation theorems. Let's stop, though, with one more and
discuss its interpretations and applications.

As a final set of example integral fluctuation theorems, we follow Ref.
\cite{Sagawa12} in defining:
\begin{align*}
i_\text{SU} \equiv \ln \frac{\Pr( \vyF , \vsF | \boldsymbol{\mu}_0 )}{ \Pr(
\vyF | \boldsymbol{\mu}_0 ) \Pr( \vsF | \boldsymbol{\mu}_0 , \vxF ) }
  ~.
\end{align*}
(This is Ref. \cite{Sagawa12}'s $I_\text{C}$, if $\boldsymbol{\mu}_0 \to \boldsymbol{\pi}_{x^0}$.)
Technically speaking, this is not a mutual information, even upon averaging. Then, we arrive at the
integral fluctuation theorems:
\begin{align*}
\left\langle  e^{-W_\text{diss} - i_\text{SU} - \Psi }
  \right\rangle_{\Pr(\vsF , \vyF , \vxF | \boldsymbol{\mu}_0 )} &= 1 
\intertext{and}
 \left\langle  e^{-W_\text{diss} - i_\text{SU} }
 \right\rangle_{\Pr(\vsF , \vyF , \vxF | \boldsymbol{\mu}_0 )} &= 1 ~.
\end{align*}
When starting from a steady-state distribution, we have the most direct
generalization of Ref.~\cite{Sagawa12}'s feedback control result, but extended to
not require detailed balance:
\begin{align*}
\left\langle  e^{-\eeep - i_\text{SU} }
  \right\rangle_{\Pr(\vsF , \vyF , \vxF | \boldsymbol{\pi_{x^0}} )} = 1
  ~.
\end{align*}
When starting from a NESS, this suggests that:
\begin{align}
\left\langle \eeep \right\rangle  \ge  -  \I_\text{SU}
  ~.
\end{align}
When the dynamics are detailed balance, this naturally reduces to the well
known results of Ref. \cite{Sagawa12} and others: $\left\langle W
\right\rangle  \ge \Delta F_\text{eq} -  \kB T \, \I_\text{SU}$.

In the feedback control setting, $Y_n$ is said to be the random variable for
measurements at time $n$. This suggests that $Y_n$ is a function of $\St_n$ and
the outcome of $Y_n$ effectively induces different Markov chains over the
states since $X_{n+1}$ is a function of $Y_n$---i.e., $x_{n+1}(y_n(s_n))$.

With our interest in complex autonomous systems, we note that our results give
new bounds on the Second Law of Thermodynamics for highly structured complex
systems strongly coupled to an environment. A preliminary application of this
was presented out in Ref.~\cite{Sartori14}. We offer our own in the next
section. Analysis of thermodynamic systems with the agency to influence their
environmental input, via some kind of coupling or feedback, say, will likely
benefit from our extended theory.

How can we reconcile this with other inequalities without auxiliary $Y$? The
other inequalities used averages of variable occurrence already conditioned on
$\vx$. However, if input $x$ and states $s$ can influence each other
dynamically through auxiliary $y$, then averaging over their joint dynamic
allows less dissipation than the traditional Second Law suggests. 

If $\St$ represents the random variable for one subset of a system's degrees of
freedom, and $Y$ represents the random variable for another subset of a
system's degrees of freedom, then the intrinsic nonextensivity of the
thermodynamic entropy $S(\St, Y | X) = S(\St | X) + S(Y | X) - \kB \I(\St; Y |
X)$ goes a long way towards explaining the physics of information stimulating
the recent resurgence of Maxwellian demonology. This viewpoint will be further
developed elsewhere.

\section{NESS Transitions in Neuronal Dynamics}
\label{sec:IonChannel}

Acting in concert, voltage-gated sodium ion channels and potassium ion channels
are the primary thermodynamic substrate that drives the evolution of membrane
potentials in neurons~\cite{Izhi10a}. Together, these voltage-gated channels are
the primary generators of the action potentials or ``spikes'' that are the
basic signals whose collective patterns support neural information
processing~\cite{Riek99}. In experiments, if the cell membrane is voltage
clamped, then the channels approach a stationary distribution over their
conformational states according to the effective energies of their biomolecular
conformations at that voltage~\cite{Hodg52}. However, absent clamping, the
channels influence their own voltage input dynamically through their current
output. The result is spontaneous spiking patterns.

Although this is not the setting in which to analyze the full richness of ion
channel interactions, we will use the sodium channel under different
voltage-driving protocols as a relatively straightforward example to
demonstrate the insights on NESS transitions gained from preceding theoretical
results. That is, while potassium ion channels are somewhat structured, the
sodium ion channel exhibits a more structured and so more illustrative dynamic
over its coarse-grained state space of functional protein conformations.

\subsection{Ion Channel Dynamics}
\label{sec:IonChannelDynamics}

\begin{figure}[htbp]
\centering
\includegraphics[width=0.5\textwidth]{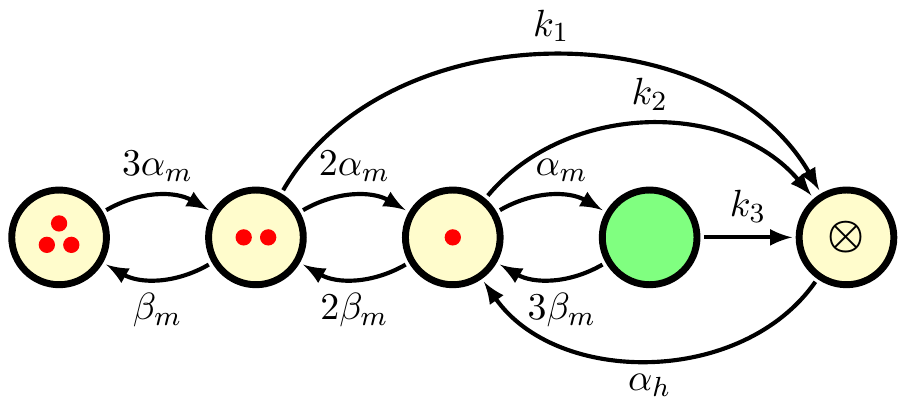}
\caption{Markov chain representation of the input-conditioned state-to-state
	rate matrix $G^{(\SSet \to \SSet | v)} $. Self-transitions are implied but,
	for tidiness, are not shown explicitly. These coarse-grainings of
	conformational states have biologically important functional
	interpretations. The number of (red) dots in each effective state
	corresponds to the number of activation gates that close off the channel.
	For example, when the channel is in the leftmost state, three activation
	gates are still active in blocking the channel. The solid (green) state
	corresponds to the channel being open. This is the only one of the five
	states in which sodium current can flow. The last state, marked
	$\boldsymbol{\otimes}$, corresponds to channel inactivation by the
	inactivation gate---when the channel is plugged by its ``ball and chain''.
	Subsequent figures use the state numbering 1 through 5 for state
	identification, which corresponds to enumeration of the states from left to
	right here.
}
\label{fig:NaChannelMM}
\end{figure}

For sodium ions to move through the neural membrane, a channel's activation
gates must be open and the deactivation gate must not yet plug the
channel~\cite{Dayan}. The rates of transitions among the conformational states
have a highly nontrivial dependence on voltage across the cell membrane. Beyond
this voltage dependence, while the activation gates act largely independently
of one another, the inactivation gate cannot plug the channel until at least
some of the activation gates are open. This causal architecture was not yet
captured by the relatively macroscopic differential equations introduced in the
pioneering work of Hodgkin and Huxley~\cite{Hodg52}. Since then, however, it
has been summarized by experimentally-motivated voltage-dependent Markov chains
over the causally relevant conformational states \cite{Patlak91}.  Here, we
follow the model implied in Ref.~\cite{Dayan}, whose voltage-dependent Markov
chain we show in Fig.~\ref{fig:NaChannelMM}.

As the mesoscopic system of thermodynamic interest, the voltage-dependent
Markov chain can be re-interpreted as a transducer that takes in voltage $v \in
\mathbb{R}$ across the cell membrane as its input and makes transitions over
its conformational state space $\SSet$ according to an infinite set of
transition matrices $\{ \T^{(\SSet \to \SSet | v)} \}_{v \in \mathbb{R}}$.
Although the set is uncountable, the voltage-conditioned transition matrices
are all described succinctly via time-independent functions of the voltage
appearing in the transition elements; denoted $\alpha_m$, $\beta_m$, and
$\alpha_h$. Time discretization of the continuous-time dynamic is
straightforward and well behaved as $\Delta t \to 0$. If the voltage $v$ is
approximately constant during the infinitesimal interval $\Delta t$, then the
state-to-state transition matrix is:
\begin{align*}
\T_{\Delta t}^{(\SSet \to \SSet | v)}
& = e^{(\Delta t) G^{(\SSet \to \SSet | v)}} \\
& \approx
I + (\Delta t) G^{(\SSet \to \SSet | v)} ~, 
\end{align*}
where $I$ is the identity matrix and $G^{(\SSet \to \SSet | v)}$ is the infinitesimal generator of time evolution:
\begin{widetext}
\begin{align}
G^{(\SSet \to \SSet | v)} \equiv
\begin{bmatrix}
- 3\alpha_m &	3\alpha_m 	&	0	&	0	&	0 	\\
\beta_m 	&	- (2\alpha_m + \beta_m + k_1) 	&	2\alpha_m &	0 &	k_1 \\
0			&	2\beta_m	& - (\alpha_m + 2\beta_m + k_2) & \alpha_m & k_2 \\
0			&	0			&	3 \beta_m 	&	- (3\beta_m + k_3) & k_3 \\
0			&	0			&	\alpha_h 	&	0	&	- \alpha_h 	
\end{bmatrix}
~.
\end{align}
\end{widetext}

\begin{figure}[htbp]
\centering
\begin{overpic}%
  [width=\linewidth] 
  {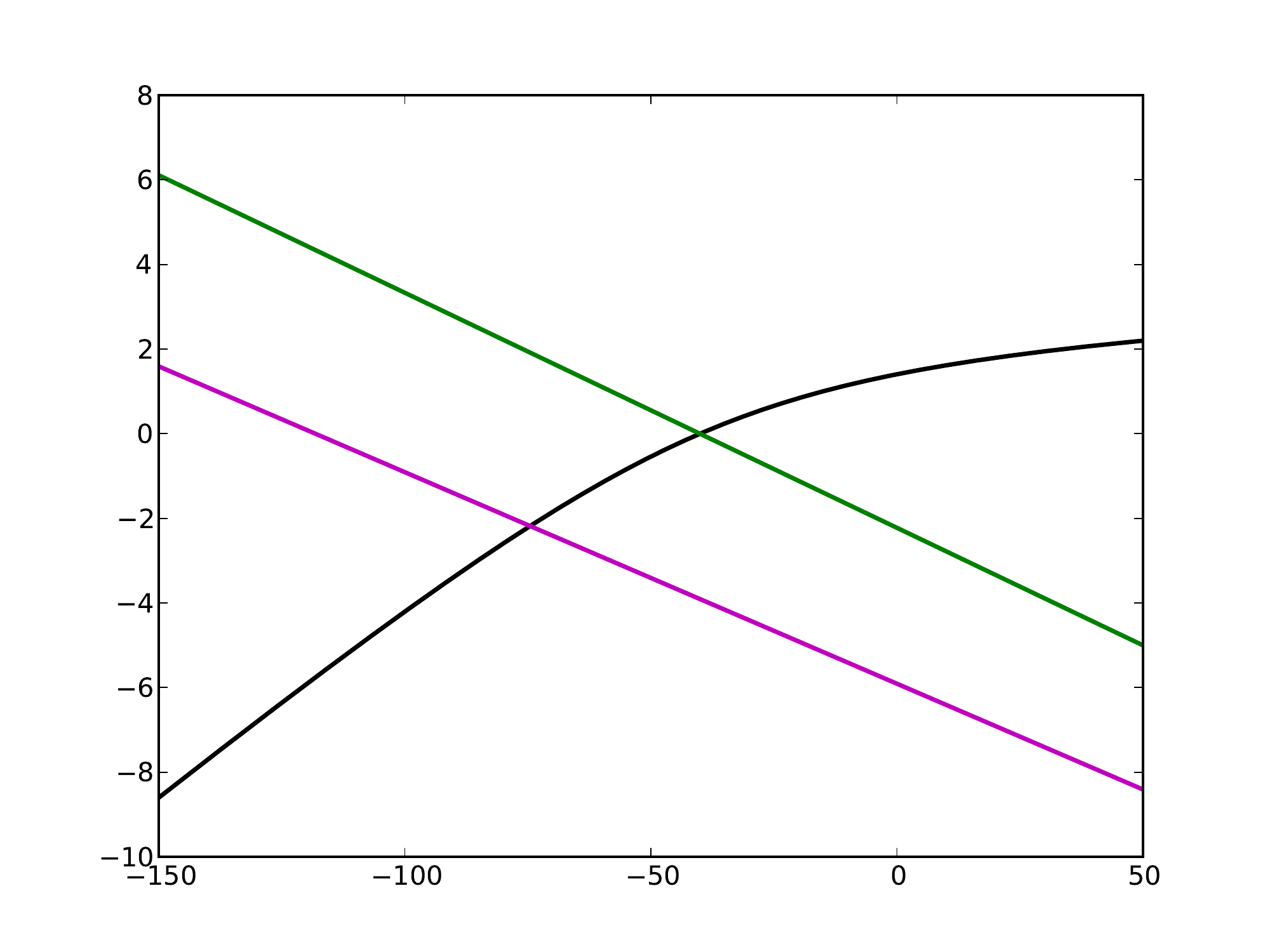}
\put(45,0){$v$ [mV]} 
\put(15,38){
    $\ln \alpha_h$}
\put(27, 55){$\ln \beta_m$}
\put(70, 50){$\ln \alpha_m$}    
\end{overpic}
\caption{Markov transition parameter voltage dependencies: Plots of
	$\ln \alpha_m$, $\ln \beta_m$, and $\ln \alpha_h$.
	These plots show that at $-100$ mV, $\beta_m
	\gg \alpha_h \gg \alpha_m \approx 0$.
	At $+10$ mV, $\alpha_m \gg \beta_m
	\gg \alpha_h \approx 0$.
}
\label{fig:ParamVDependence}
\end{figure}

Specifically, $\alpha_m$, $\beta_m$, and $\alpha_h$ are voltage-dependent
variables, as found in the Hodgkin and Huxley model \cite{Hodg52, Dayan}:
\begin{align*}
\alpha_m(v) & = \frac{(v + 40 \text{ mV}) / 10 \text{ mV}}{1 - \exp \left[ -(v + 40 \text{ mV}) / 10 \text{ mV} \right]} ~, \\
\beta_m(v) & = 4 \exp \left[ -(v + 65 \text{ mV}) / 18 \text{ mV} \right] \phantom{\Biggl( \Biggr)}
  ~,
\end{align*}
and:
\begin{align*}
\alpha_h(v)
  & = \tfrac{7}{100} \exp \left[ -(v + 65 \text{ mV}) / 20 \text{ mV} \right]
  ~.
\end{align*}
See Fig.~\ref{fig:ParamVDependence}.
The reaction-rate constants are $k_1 = 6/25 \text{ ms}^{-1}$, $k_2 = 2/5 \text{
ms}^{-1}$, and $k_3 = 3/2 \text{ ms}^{-1}$.

We developed new spectral decomposition methods from the meromorphic functional
calculus \cite{Crut13a,Riec16a} to circumvent the inherent ill-conditioning in
ion channel dynamics. Using these we can analytically calculate most, if not
all, properties---e.g., dynamics, expected current, thermodynamics, information
measures, and the like---about this model directly from the transition dynamic.

\begin{figure}[htbp]
\centering
\begin{overpic}%
  [width=\linewidth]
  {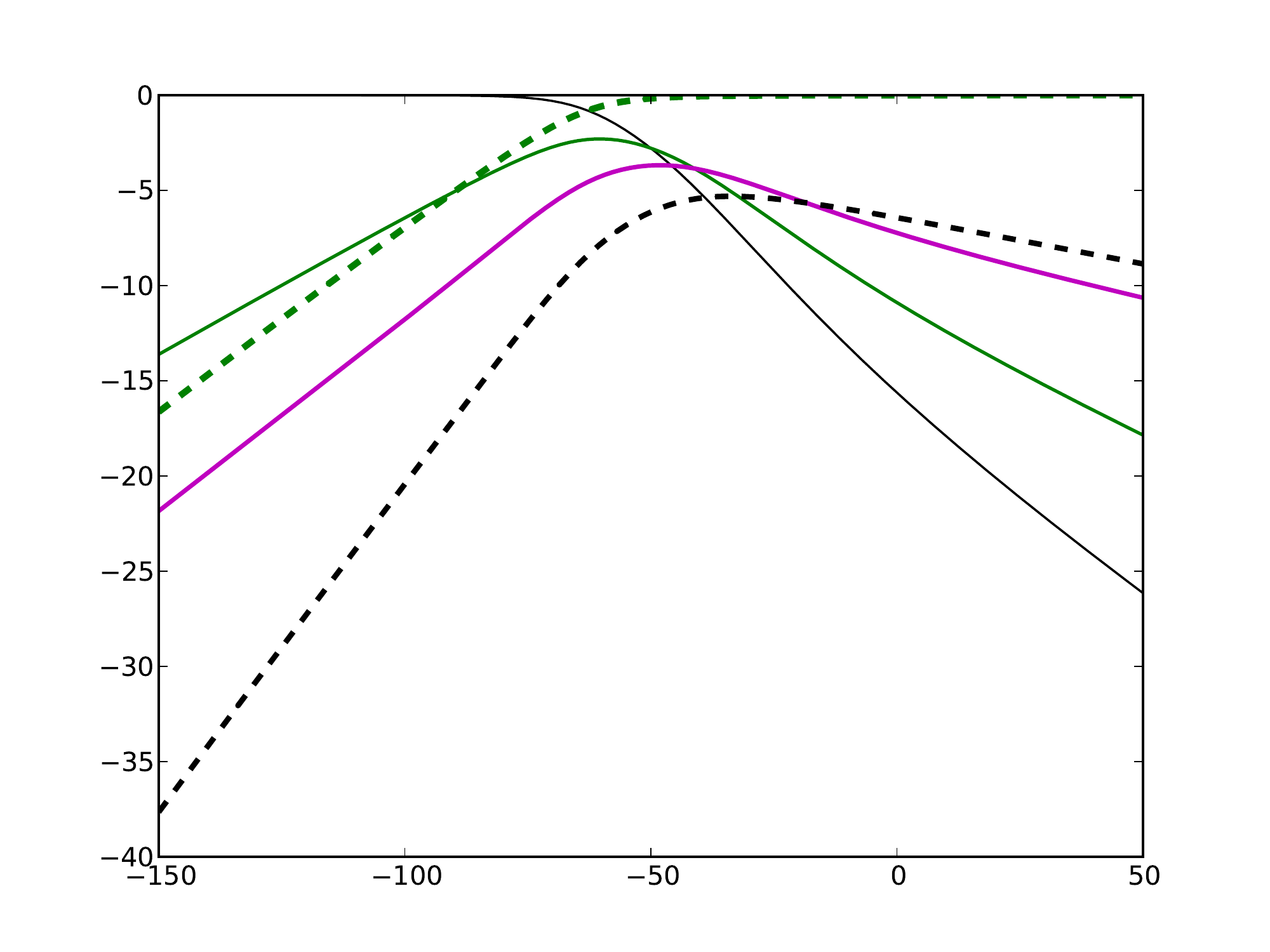}
\put(45,0){$v$ [mV]} 
\put(19,63){1}
\put(19,53){2}
\put(19,35){3}
\put(22,20){4}
\put(80,63){5}
\end{overpic}
\caption{Steady-state distribution voltage dependence: Negative of the
	steady-state surprisals, $\ln \pi_v(s) = -\phi(v,s)$ for each
	conformational state. Each curve labeled by the state-number to which it
	corresponds. Note that $-100$ mV and $+10$ mV (relevant for later) are
	extremes in that $\boldsymbol{\pi_{\va}} \approx \boldsymbol{\delta_1}$ and
	$\boldsymbol{\pi_{\vb}} \approx \boldsymbol{\delta_5}$.
}
\label{fig:NESSsurprisals}
\end{figure}

Since we are interested in thermodynamics, though, let us focus on determining
the steady-state surprisals of the conformational states. For any persistent
environmental input, the effective energies of the various conformational
states are determined by their relative stationary occupation probability;
according to Eq.~\eqref{eq:NEPdef}, $\pi_{x}(s) = e^{- \phi (x,s) }$. The
stationary distribution $\boldsymbol{\pi_v }$ induced by persistent $v$ is the
left eigenvector of $T_\tau^{(\SSet \to \SSet | v)}$ associated with the
eigenvalue of unity. Equivalently, and more convenient in this case,
$\boldsymbol{\pi_v }$ is the left eigenvector of $G^{(\SSet \to \SSet | v)}$
associated with the eigenvalue of zero. Via $\bra{\boldsymbol{\pi_v } }
G^{(\SSet \to \SSet | v )} = \vec{0}$, we find that the steady state
distribution for any persistent $v$ is:
\begin{align*}
\boldsymbol{\pi_v } & \propto \Bigl( \frac{\beta_m}{3 \alpha_m} \, , \, \; 1 \, , \, \; \frac{2 \alpha_m + k_1}{2 \beta_m} \, , \, \; 
\frac{\alpha_m}{2 \beta_m} \left(\frac{2 \alpha_m + k_1}{3 \beta_m + k_3}
\right) \, , \, \;  \\
& \qquad \qquad \frac{1}{\alpha_h} \Bigl[ k_1 + \frac{2 \alpha_m + k_1}{2 \beta_m} \left( k_2 + \frac{k_3 \alpha_m}{3 \beta_m + k_3} \right) \Bigr] \Bigr) \\
& \propto e^{- \boldsymbol{\phi}(v) }
~,
\end{align*}
which immediately yields the steady-state surprisals for conformational states
at a constant environmental input $v$. The steady-state surprisal is shown for
each conformational state in Fig.~\ref{fig:NESSsurprisals}, as a function of
the voltage-clamped membrane potential $v$.

\begin{figure}[htbp]
\centering
\begin{overpic}%
  [width=\linewidth]
  {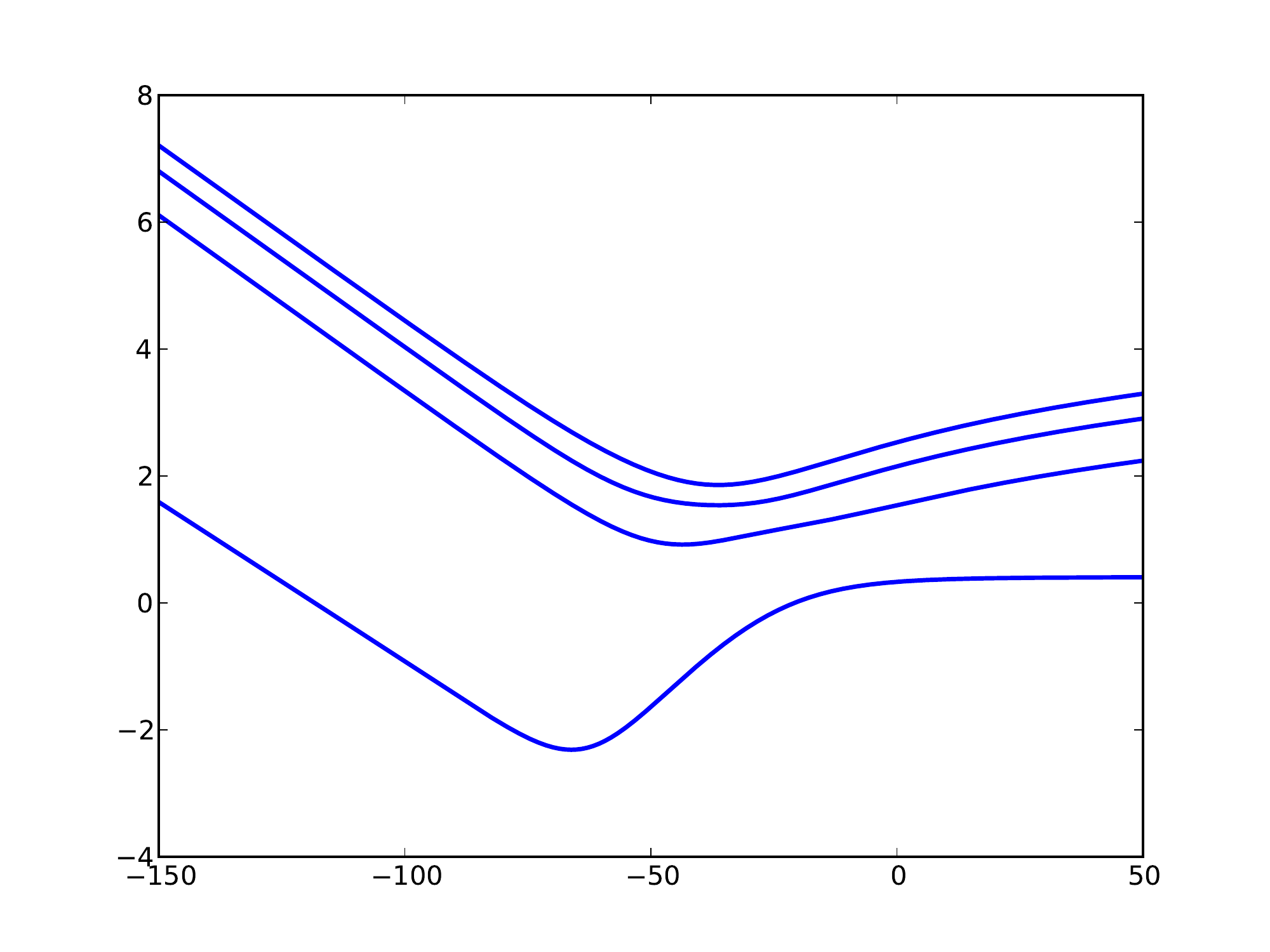}
\put(45,0){$v$ [mV]} 
\end{overpic}
\caption{Modes of the state-to-state dynamic: All eigenvalues of $G$ are real
	and nonpositive. There is a zero eigenvalue associated with stationarity
	and four negative eigenvalues associated with decay rates from the states.
	Plots of $\ln [ - \lambda(v) ]$ for $\lambda(v) \in \Lambda_{G^{(\SSet \to
	\SSet | v)}}$. Smaller $\ln(- \lambda)$ corresponds to longer time-scales.
	The zero eigenvalue maps to $- \infty$.
	}
\label{fig:Geigs}
\end{figure}

\subsection{Ion channel (ir)reversibility}

Recall that detailed balance is the condition that, for states $a$ and $b$ and
environmental input $x$, $\Pr(b \xrightarrow{x} a) / \Pr(a \xrightarrow{x} b) =
\pi_x(a) / \pi_x(b)$.  Interestingly, in this biologically inspired model,
detailed balance is satisfied by \emph{several, but not all} of the
state-transition pairs.

For example, for very small $\Delta t$:
\begin{align*}
\frac{\Pr(1 \xrightarrow{v} 2)}{\Pr(2 \xrightarrow{v} 1)}
  & = \frac{3 \alpha_m}{\beta_m} \\
  & =  \frac{\pi_v(2)}{\pi_v(1)}
  ~.
\end{align*}
That is, this transition pair satisfies detailed balance. 
Hence, all transitions between these states are completely reversible:
$\Psi(2 | 1, v) = \Psi(1 | 2, v) = \Psi(22121112 | 2, v) = 0$.

However, this does not hold for other transition pairs. Consider
transitions between states $2$ and $3$:
\begin{align*}
\frac{\Pr(2 \xrightarrow{v} 3)}{\Pr(3 \xrightarrow{v} 2)} 
  & = \frac{\alpha_m}{\beta_m}  \\
  & \neq  \frac{\pi_v(3)}{\pi_v(2)} \\
  & = \frac{\alpha_m + k_1 / 2}{\beta_m}
  ~.
\end{align*}
Most other transitions also violate detailed balance.

Since detailed balance does not hold for the effective dynamic, the theory
developed above is essential to analyzing the sodium ion channel
thermodynamics. Moreover, the fact that $G^{(\SSet \to \SSet | v)}$ has null
entries that are nonzero for its transpose implies that paths involving these
transitions will be infinitely irreversible: $\Psi = \infty$ for such paths as
$\Delta t \to 0$; specifically, the transitions of $G^{(\SSet \to \SSet | v)}$
with rates $k_1$ and $k_3$. Forbidden transitions are an extreme form of
irreversibility that are nevertheless commonly observed for complex systems, as
the ion channel so readily illustrates. In it, the asymmetry in allowed
transitions can be traced to different \emph{mechanisms} facilitating different
paths through the state space. Whether the irreversibility is truly infinite or
just practically infinite does not matter much for the excess thermodynamics,
although it will of course affect the calculated distribution of $\Psi$.
(Conventional, linear algebraic methods are inadequate to overcome these
technical challenges. The spectral decomposition methods, mentioned above, are
required.)

\subsection{Step Function Drive}

With this understanding of ion channel NESSs, let's now turn to the
thermodynamics induced by driving between them. We first consider the
particular voltage protocol of $\va \equiv -100$ mV for all time except a $\vb
\equiv 10$ mV pulse for $5$ ms starting at $t = 0$. This is an example of
continuous-time dynamics and deterministic driving. The system begins
equilibrated with the static environmental drive $\va =- 100$ mV. The initial
distribution over $\SSet$ is thus $\boldsymbol{\mu}_0 = \boldsymbol{\pi_{\va}
}$, where $\boldsymbol{\pi_{\va} }$ is the left eigenvector of $G^{(\SSet \to
\SSet | \va)}$ associated with the eigenvalue of zero.

During an epoch of fixed $v = V$, the net transition dynamic after $\tau$ ms becomes: 
\begin{align*}
T_\tau^{(\SSet \to \SSet | v = V)} = e^{\tau G^{(\SSet \to \SSet | v = V)}}
  ~.
\end{align*}
Therefore, the distribution over states induced by the driving protocol is:
\begin{align}
\bra{ \boldsymbol{\mu}_t } & =
\begin{cases}
\bra{\boldsymbol{\pi_{\va} }} & \text{for } t \leq 0 \\
\bra{\boldsymbol{\pi_{\va} }}   e^{t G_\text{b} } & \text{for } 0 < t \leq 5 \text{ ms} \\
\bra{\boldsymbol{\pi_{\va} }} e^{5 G_\text{b} } e^{(t-5) G_\text{a} } & \text{for }  t > 5 \text{ ms} 
\end{cases} ~,
\label{eq:ExDistr}
\end{align}
where, for brevity, we defined: 
$G_\text{a} \equiv G^{(\SSet \to \SSet | \va)}$ and 
$G_\text{b} \equiv G^{(\SSet \to \SSet | \vb)}$.
These are especially useful when expressing the rate matrix via its spectral
decomposition, using the methods of Refs. \cite{Crut13a,Riec16a}. Besides the
zero eigenvalue, there are only four other eigenvalues of $G$ that are
determined via det$(\lambda I - G) = 0$.

\begin{figure}[htbp]
\centering
\begin{overpic}%
  [width=\linewidth]
  {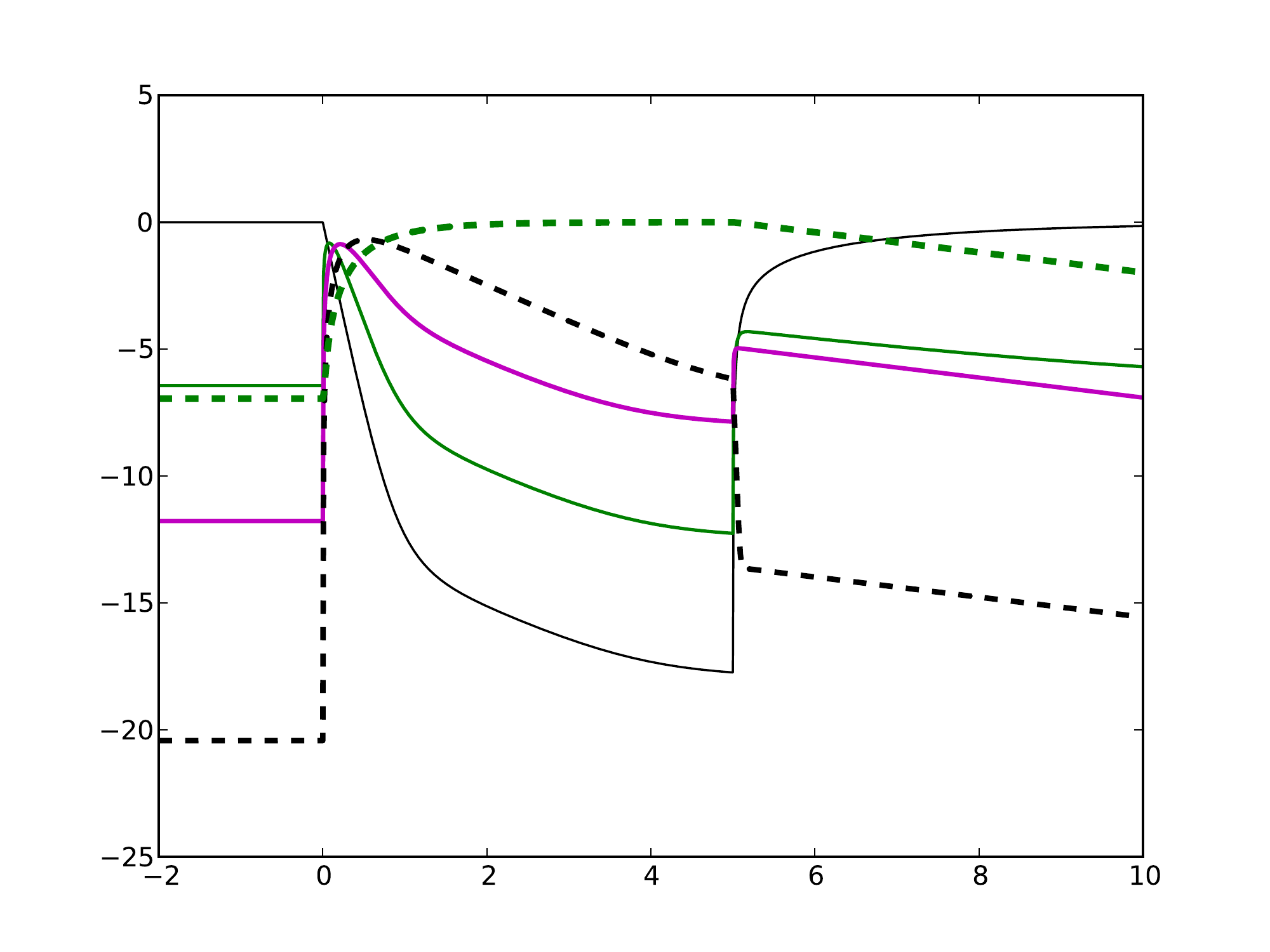}
\put(45,0){$t$ [ms]} 
\put(19,60){1}
\put(18,48){2}
\put(18,35){3}
\put(21,20){4}
\put(45,60){5}
\end{overpic}
\caption{Na$^+$ ion channel NESS transitions: temporal evolution of the
	distribution $\boldsymbol{\mu}_t$ of ion-channel conformational states
	induced by a deterministic $5$ ms voltage pulse is shown via plots of $\ln
	\braket{ \boldsymbol{\mu}_t | s}$ for all $s \in \SSet$. Curves are labeled
	by the state-number to which each component corresponds.
}
\label{fig:mu4detVpulse}
\end{figure}
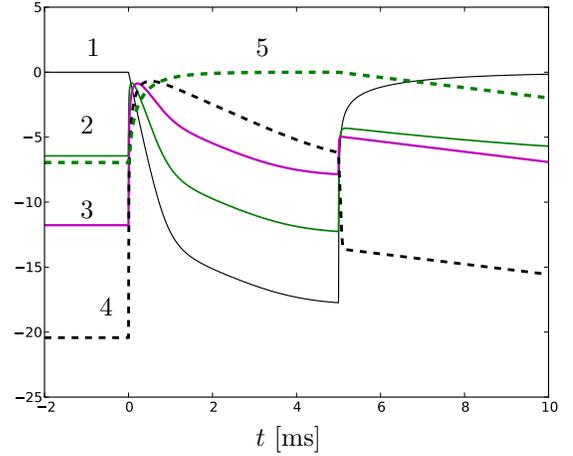

Figure~\ref{fig:Geigs} shows $G$'s eigenvalues as a function of $v$,
which indicates the voltage-dependent timescales of probability decay for modes of occupation probability. 
The associated decay rates play a prominent role in
Fig.~\ref{fig:mu4detVpulse}, which shows the time-dependent distribution induced over
states by the 5 ms voltage pulse driving protocol.

Having the distribution over states at all times is powerful knowledge.  For
example, since the current through a single channel is binary---either $0$ or
$I_0(v)$---and since current only flows in the open conformation, the expected
current through the channel is $I_0(v) = g_0 [ v-V_\text{Na} ]$ times the
expectation value of being in the open state:
\begin{align*}
\left\langle I (t) \right\rangle 
& = g_0 \left[ v(t) - V_\text{Na}  \right]  \braket{ \boldsymbol{\mu}(t) | \delta_{\text{open}}} ~,
\end{align*} 
where $\delta_{\text{open}} = (0, 0, 0, 1, 0)$, $g_0$ is the conductance of an open Na$^+$ channel, and $V_\text{Na} = \frac{\kB T}{e^+} \ln
\frac{[\text{Na}^+]_\text{out}}{[\text{Na}^+]_\text{in}} \approx 90$ mV is the
Nernst potential for sodium in a typical mammalian neuron~\cite{Dayan}. To be
clear $\left\langle I (t) \right\rangle$ is what would be observed from an
ensemble of channels in a local patch of cell membrane experiencing the same
driving. The current produced from the Markovian model appears to be more
realistic than what would be expected from the Hodgkin--Huxley
model~\cite{Dayan}. Moreover, using our spectral-decomposition methods for
functions of a Markov chain \cite{Crut13a,Riec16a}, this current can now be
obtained in closed-form.

Let us start the thermodynamic investigation by considering excess work $\Wex$.
With $\tau = N \Delta t$, we take the limit of $\Delta t \to 0$ while keeping
the product $N \Delta t = \tau$ constant. 
Then the expected excess work per $\kB T$, from time $t_0$ to time $t_0 + \tau$, is:
\begin{align*}
\beta \left\langle  \Wex \right\rangle
  & = \int_{t_0}^{t_0+\tau} \braket{ \boldsymbol{\mu}_t | d \phi_{v(t)}/dt } \, dt
  ~.
\end{align*}
However, it should be clear that, for this stepped voltage protocol, excess
work is \emph{only performed on this system at the very onset and subsequently
at the end} of the step driving. Indeed, this is the only time that the driving
$v(t)$ changes and, thus, the only time that the state-dependent rate of work
$d \ket{\phi_{v(t)}}/dt$ is nonzero. As we let $\Delta t \to 0$, the expected excess work (divided by $\kB T$) near the onset of driving becomes a step function with height:
\begin{align*}
\lim_{\epsilon \to 0^+}
  & \langle \eeep (t=\epsilon) - \eeep (t=-\epsilon) \rangle  \\
  & = \sum_{s \in \SSet} \braket{\boldsymbol{\pi_{\va} } | s }
  \bigl[ \phi(10 \text{mV}, s) - \phi (-100 \text{mV}, s) \bigr]
  ~,
\end{align*}
where $\braket{\boldsymbol{\pi_{\va} } | s } = \pi_{-100 \text{ mV}}(s)$.

Indeed, for this singular event, the full distribution of work performed can be
given according to the probabilities that the system was in a particular state
when the driving was applied. For $0 < t < 5$ ms, the probability density
function for $\beta \Wex$ is:
\begin{align*}
p(\eeep) & = \sum_{s \in \SSet} \pi_{-100 \text{ mV}}(s) \\
	& \qquad \times \delta
	\Bigl(
	\eeep - \bigl[ \phi(10 \text{mV}, s) - \phi (-100 \text{mV}, s) \bigr]
	\Bigr)
  ~,
\end{align*}
where $\delta(\cdot)$ here is the Dirac delta function. For $t > 5$ ms, the
full excess environmental entropy production (EEEP) probability density
function (pdf) is:
\begin{align*}
p(\eeep) & = \sum_{s, s' \in \SSet}
  \braket{\boldsymbol{\pi_{\va}} | s } \bra{ s } e^{5 G_\text{b} } \ket{ s' } \\
  & {\small{ \times \delta \Bigl( \eeep - \bigl[ \phi(\vb, s) - \phi (\va, s) \bigr] - \bigl[ \phi(\va, s') - \phi (\vb, s') \bigr] \Bigr) }}
  .
\end{align*}
From the Dirac delta function's argument and the sum over $s$ and $s'$, it is
clear that every nonzero-probability EEEP value $\eeep$ also has a nonzero
probability for the negative $-\eeep$ of that EEEP value.

\begin{figure}[htbp]
\centering
\begin{overpic}[width=\linewidth]{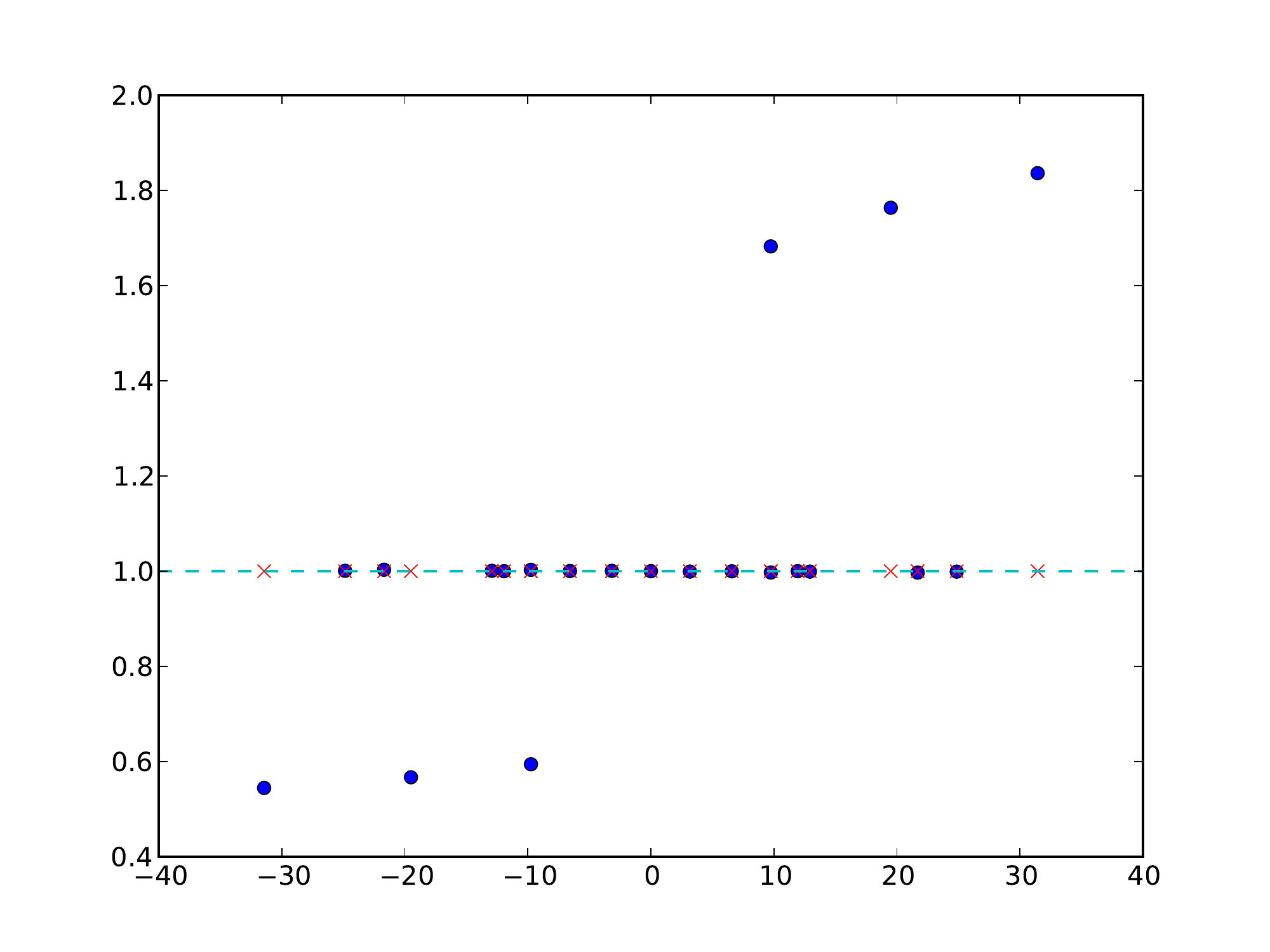}
\put(0,38){\rotatebox[origin=c]{90}{$\frac{\Pr(\eeep)}{\Pr(-\eeep)} / e^\eeep$}}  
\put(49,0){$\eeep$}
\end{overpic}
\caption{Deviations from and agreement with the Crooks Fluctuation Theorem
	for nondetailed balance dynamics: Exact calculation of
	$\frac{\Pr(\eeep)}{\Pr(-\eeep)} / e^\eeep$ (blue dots) for all allowed
	values of $\eeep = \beta \Wex$ during the pulse drive. Since the system
	starts in equilibrium and since the driving is time-symmetric, a naive CFT
	interpretation suggests that all values lie at unity, dashed line (blue)
	and marked with a $\times$ (red) wherever an allowed excess work value
	appears. Interestingly, many of the allowed work values \emph{do} still
	fall on or very near unity. Absent detailed balance, though,
	Eq.~\eqref{eq:genCFT} must be used to account for the actual distribution
	of excess environmental entropy production and path irreversibilities that,
	in addition to all other values, yields the six deviant markings (blue
	dots) above and below unity.
}
\label{fig:PrWvsPrmW}
\end{figure}

For the time-symmetric 5 ms voltage-pulse driving, Eq.~\eqref{eq:TSgenCFTwW}
tells us that $\Pr(\eeep , \Psi ) / \Pr(- \eeep , - \Psi ) = e^\Psi e^\eeep$.
Since there are infinitely many $\Psi$ values to account for, we do not plot
the joint distribution explicitly. However, we can appreciate the necessity of
the relationship by comparing it to the naive CFT interpretation that, for this
case, suggests $\Pr(\eeep ) / \Pr(- \eeep) = e^\eeep$.
Figure~\ref{fig:PrWvsPrmW} compares these by plotting $e^{- \eeep} \, \Pr(\eeep
) / \Pr(- \eeep) $.

Allowed values of the excess work that do \emph{not} lie on
$e^{- \eeep} \, \Pr(\eeep ) / \Pr(- \eeep) = 1$ demonstrate deviations from the
naive CFT interpretation. Since the constant-voltage steady states are
nonequilibrium and, thus, not microscopically reversible---i.e., $\Psi \neq 0$
for some state paths---the naive CFT interpretation cannot be true despite the
time-symmetric driving. Perhaps the most surprising feature in
Fig.~\ref{fig:PrWvsPrmW} is that many of the probability ratios still \emph{do}
(almost) fall on the naive CFT line at unity. In part, this is due to a subset
of the cycles in the NESS dynamic obeying detailed balance. Another
contributing factor is that longer durations $\tau$ of fixed $v$ induces a
\emph{net} dynamic $e^{\tau G}$ that \emph{approaches} a detailed-balanced
dynamic. That the values in Fig.~\ref{fig:PrWvsPrmW} are sensible can be
verified by checking the ratio of the joint probabilities
$\braket{\boldsymbol{\pi_{\va}} | s } \bra{ s } e^{5 G_\text{b} } \ket{ s' }$
to the value of the joint probability with $s$ and $s'$ swapped.

In stark contrast to the instantaneous work contribution, the system's excess
heat $\Qex$ unfolds over time, exhibiting a rich structure governed by the
trajectories through the conformational state-space. The expected excess heat
per $\kB T$ is:
\begin{align}
\beta \braket{\Qex}
& = \int_{t_0}^{t_0+\tau} \braket{\dot{\boldsymbol{\mu}} | \phi_{v(t)} } \, dt ~. 
\label{eq:HeatIntegral}
\end{align}
over a duration $\tau$, if starting at time $t_0$. Since $v(t)$ is constant
except at the two instants of change, the integral is easily solved exactly
using the fundamental theorem of calculus and Eq.~\eqref{eq:ExDistr}.

\begin{figure}[htbp]
\centering
\begin{overpic}%
  [width=\linewidth] 
  {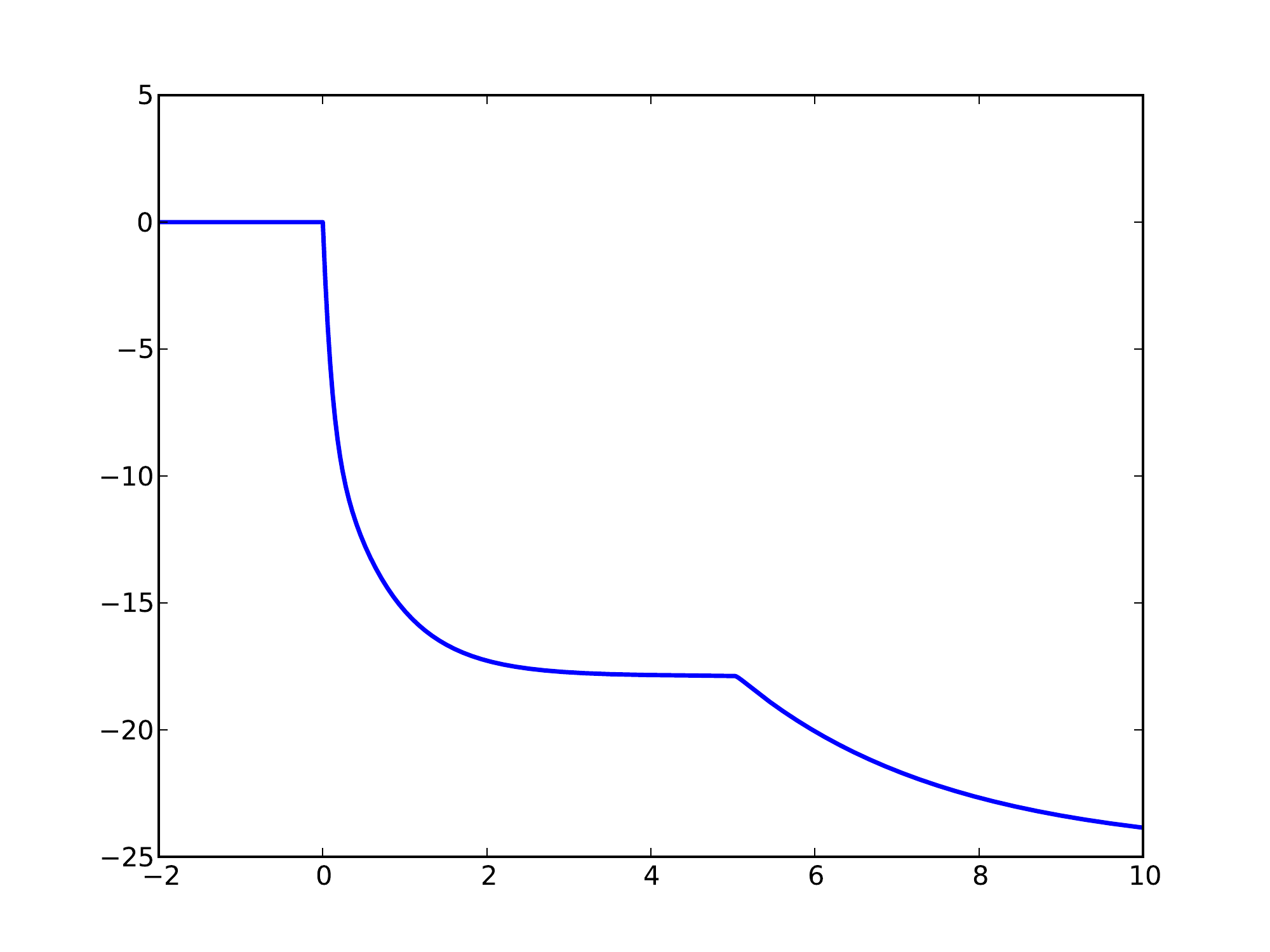}
\put(45,0){$t$ [ms]} 
\put(3, 35){\rotatebox[origin=c]{90}{$\braket{\Qex} \; [\kB T]$}}  
\end{overpic}
\caption{Excess heat $\Qex$ per ion channel for an ensemble of ion channels embedded in a local patch of cell membrane;
in units of $\kB T \approx 26$ meV.
The two bouts of relaxation correspond to the ion channel adapting
to sudden changes in voltage across the cell membrane.
	}
\label{fig:Heat4det5msPulse}
\end{figure}

For $t_0 < t_0 + \tau < 0$, the system is in the initial steady-state and has a
time-independent heat and so a constant excess heat rate that vanishes:
\begin{align*}
\frac{\braket{\Qex} }{\tau}
  & = 0
  ~.  
\end{align*}

Figure~\ref{fig:Heat4det5msPulse} shows the expected excess heat $\langle \Qex
\rangle$ over the course of the voltage-drive protocol. The steady-state
average rate of excess heat production within any steady state is necessarily
zero. However, the channel macromolecule responds to changes in the environment
via conformational changes and corresponding heat productions that unfold on
the timescale of milliseconds \footnote{The characteristic timescale is
actually the net result of a combination of timescales from the inverse
eigenvalues of $G$. Of necessity, these are the same timescales that determine
the relaxation of the state distribution.}. Notably, the expected heat is on
the order of tens of $\kB T$, which for mammalian neurons is $\kB T =
1/\beta \approx 26$ meV.

For $0 = t_0 < t_0 + \tau < 5$ ms, the system has a time-varying heat
production as it synchronizes to the new nonequilibrium steady state. From
Eq.~\eqref{eq:ExDistr}, we know that $\bra{ \boldsymbol{\mu}_t } = \bra{
\boldsymbol{\pi}_{\va} }  e^{t G_\text{b} } $ in this case. At the same time,
the steady-state-surprisal $\ket{\phi_{v(t)} }$ is time-independent during this
epoch since $v$ is temporarily fixed at $\vb = 10$ mV.

Again, as seen in Fig.~\ref{fig:Heat4det5msPulse}, the expected excess heat
drops for several milliseconds after the final voltage switch at 5 ms, as the
ion channel re-adapts to its original steady state. For this second bout of
relaxation the adaptation is slower since, in accordance with
Fig.~\ref{fig:Geigs}, the slowest timescale at $\va = -100$ mV is slower than
the slowest timescale at $\vb = 10$ mV.

Overall, we see that the excess thermodynamic quantities are well behaved and
accessible: without needing to know the background biological upkeep of the
Na$^+$ ion channel, we can access and control coarse degrees of freedom of the
channel macromolecule via modulating the voltage across the cell membrane.
Moreover, for the Na$^+$ channel, state-measurement and feedback on the
timescale of milliseconds would allow significant alterations of heat and
entropy production. Fortuitously, this suggests an accessible platform for
laboratory experimentation. Next, we comment briefly on an intrinsic type of
measurement and feedback that happens in vivo every moment.

\subsection{Intrinsic feedback}

Having come this far, we close illustrating the thermodynamics of NESS
transitions with a final application. In a biologically active (in vivo)
neuron, the input membrane voltage at each time depends on integrated current---a
functional of the state distribution---up to that time. Our relations for
modified integral fluctuation relations describe the thermodynamic agency of
Na$^+$ channels in vivo, whereas conventional fluctuation relations fall
short.  Although there is certainly feedback in vivo, it is not the ``feedback
control'' discussed recently. Importantly, no ``outsider'' forces the feedback;
the feedback is intrinsic---woven into the system--environment joint dynamic.
We leave a thorough investigation of the thermodynamics of intrinsic feedback
to elsewhere. The success here, however, already suggests investigating other
natural systems with intrinsic feedback---in joint nonlinear dynamics and
complex networks---to test the new fluctuation theorems and computational
methods in a broader class of interacting complex nonequilbrium systems.

\section{Discussion}
\label{sec:Discussion}

In light of our refined detailed fluctuation theorem Eq.~\eqref{eq:genCFT} for nondetailed-balanced dynamics, we referred to the common belief in
Eq.~\eqref{eq:CFTviolation} as the naive CFT interpretation since it appeals to
a nonphysical conjugate dynamics, as in Eq.~\eqref{eq:StandardCFT}. Similarly,
we referred to failures of Eq.~\eqref{eq:StandardCFTInterpreted} as CFT
violations. Nonetheless, with proper interpretation using the unphysical
conjugate dynamics, Eq.~\eqref{eq:StandardCFT} is mathematically correct even
without detailed balance and can be a useful device for establishing integral
FTs.

The CFT is often misinterpreted, though, despite receiving widespread
attention. For example, Ref.~\cite{England13} appealed to the naive CFT
interpretation in the case of nondetailed balance dynamics of self-replication.
However, as we showed, such applications are not valid. So, the statistical
physics of self-replication either depends on an assumption of detailed balance
or deserves a generalization. We believe the latter should be straightforward
using our results, new spectral methods, and a derivation paralleling Ref.~\cite{England13}.

We hope that our nonintegral FTs---especially Eq.~\eqref{eq:genCFT} that
constrains the joint distribution of excess and housekeeping entropies---will
provide better physical intuition for the structure of effective dynamics
outside detailed balance. Path irreversibility clearly plays a prominent role.
Although the preceding introduced a unifying framework, certain subclasses of
path irreversibility have already been proposed recently.

In certain applications, for example, path irreversibility is governed by
differences in chemical potential. In such cases, the irreversibility is
quantitatively related to cycle affinities; see, for example, Ref.
\cite{Lacoste08}. It essentially discovered a special case of the results
developed here specifically applicable to the interesting example of a kinesin
motor protein.

Recently, Ref.~\cite{Murashita14} elaborated on one type of irreversibility,
called \emph{absolute irreversibility}, that at first appears to constitute an
extreme contribution to the total path irreversibility $\Psi$. This indeed is
one interpretation, but not the full story. On closer examination, its result
appears to coincide most directly with Eq.~\eqref{eq:WdissIFT} which must be
used when starting in a nonsteady-state, rather than with a violation of
Eq.~\eqref{eq:BasicIFT} which is simply inapplicable when starting in a
nonsteady-state. We reinterpret that work as a testament to the importance of
the nonsteady-state contribution to free energy change, $\beta^{-1} \Delta
(\gamma)$. Explicitly: 
\begin{align*}
\left\langle  e^{-\beta W_\text{diss}}
  \right\rangle_{\Pr( s_{0:N} | \StartF, \xF) } & = 1 \\
  \to \braket{\eeep} & \geq \Delta \braket{ \gamma } ~.
\end{align*}
This captures, for example, the entropy change associated with free expansion.
From our viewpoint, however, any absolute irreversibility is only one extreme
of the broader generalization introduced above to explore the consequences of
irreversibility and nonsteady-state additions to free energy.

To frame our results in yet another way, we note that the ``feedback control''
imposed by an experimenter on an otherwise detailed-balanced system is a rather
limited form of CFT violation. Yet it appears to be the only one having gained
much recognition. This is odd. Hysteresis, to take one example, common in
paradigmatically complex physical systems, provides a more physical
manifestation of CFT violation. Even this is still a relatively tame deviation
from detailed balance. Living systems are the true flagship of complex physical
agents with intrinsic computational feedback across many levels of their
organization. Our fluctuation relations describe all of these aspects, together.   

In particular, they suggest how a system's intrinsic model of
its environment, together with an action policy that leverages knowledge
captured in the model to control the environment, allows the system to play the
survival game to its thermodynamic advantage. For example, an agent can use
information about the environment to increase its nonsteady-state free energy
and perform useful work---a phenomenon that is not only reminiscent of living
beings, but also comes very near to defining them.

We hope that our results and methods stimulate investigating the excess
thermodynamics of systems with intrinsic feedback---from designed ``toy
demons'' to complex biological molecules affected by and simultaneously
affecting their environments. Several biological examples that suggest
themselves include kinesin motors~\cite{Altaner15}, drug-operated
channels~\cite{Colquhoun77}, and dynamic synapses~\cite{Lahiri13}, just to name
a few.

\section{Conclusion}

We presented generalized fluctuation theorems for complex nonequilibrium
systems driven between NESSs. In addition to the detailed FTs that constrain
joint distributions of excess and housekeeping quantities, we introduced
integral fluctuation theorems in the presence of an auxiliary variable. The
auxiliary variable need not be measurement nor any other meddling of an
outsider. Due to this, it generalizes the theory of ``feedback control'' to the
setting of arbitrary intrinsic feedback between system and environment.

A sequel to the above derives exact closed-form expressions for the moments of
excess heat and excess work when the joint system--environment dynamic is
governed by a (finite or countably infinite) discrete- or continuous-time
hidden Markov model. A joint system can always be modeled as a joint hidden
Markov model---at least as an approximation to the true joint dynamics. For
this reason, our exact results should provide broadly applicable tools. The
latter have particular theoretical advantage in giving access to what occurs in
transient and asymptotic dynamics of excess thermodynamic quantities atop
NESSs.

In summary, the traditional laws of thermodynamics are largely preserved for
the renormalized ``excess'' thermodynamic quantities that arise naturally when
considering nondetailed-balanced complex systems. However, the laws must be
modified by the entropic contribution of path irreversibility. We noted that
the latter turns out to be equivalent to steady-state thermodynamics'
housekeeping entropy.

Our relations still hold for excursions between equilibrium steady states, but
we then have the simplification that $\Psi = \beta Q_\text{hk} = 0$.
Consistently, equilibrium thermodynamics is a reduction of the theory of excess
thermodynamic quantities with no housekeeping terms---when all paths are
microscopically reversible.

Layers of emergence, typical of the biological world \cite[Fig. 6]{Shen06a},
beg renormalization in terms of a hierarchy of housekeeping backgrounds
\cite{Anderson72}. The opportunity offered up by emergent levels of novel
organization is a new richness in nondetailed-balanced effective
dynamics---dynamics and structure that can be exploited by intelligent
thermodynamic agency \cite{Boyd16c,Boyd16d}. We consider the thermodynamics of
agency in a sequel, analyzing a simple autonomous agent that harvests energy by
leveraging hidden correlations in a fluctuating environment.

We leave the development for now, but with an encouraging lesson: Even in
nonstationary nonequilibrium, there is excess thermodynamic structure 
at any level of observation
that we can access, control, and harness.

\section*{Acknowledgments}

We thank Tony Bell, Alec Boyd, Gavin Crooks, Chris Jarzynski, John Mahoney,
Dibyendu Mandal, and Adam Rupe for useful feedback. We thank the Santa Fe
Institute for its hospitality during visits. JPC is an SFI External Faculty
member. This material is based upon work supported by, or in part by, the U. S.
Army Research Laboratory and the U. S. Army Research Office under contracts
W911NF-12-1-0234, W911NF-13-1-0390, and W911NF-13-1-0340.

\appendix

\section{Extension to Non-Markovian Instantaneous Dynamics} 
\label{sec:HMMs}

Commonly, theoretical developments assume state-to-state transitions are
instantaneously Markovian given the input. This assumption works well for many
cases, but fails in others with strong coupling between system and environment.
Fortunately, we can straightforwardly generalize the results of stochastic
thermodynamics by considering a system's observable states to be functions of
latent variables $\MxSSet$. The goal in the following is to highlight the
necessary changes, so that it should be relatively direct to adapt our
derivations to the non-Markovian dynamics.

\subsubsection{Latent states, system states, and their many distributions}

Even with constant environmental input, the dynamic over a system's states need
not obey detailed balance nor exhibit any finite Markov order. We assume that
the classical observed states $\SSet$ are functions $f: \MxSSet \to \SSet$ of a
latent Markov chain. We also assume that the stochastic transitions among
latent states are determined by the current environmental input $x \in
\mathcal{X}$, which can depend arbitrarily on all previous input and
system-state history. The Perron--Frobenius theorem guarantees that there is a
stationary distribution over latent states associated with each fixed input
$x$; the function of the Markov chain maps this stationary distribution over
latent states into the stationary distribution over system states. These are
the stationary distributions associated with system NESSs.

We assume too that the $\MxSSet$-to-$\MxSSet$ transitions are Markovian given
the input. However, different inputs induce different Markov chains over the
latent states. This can be described by a (possibly infinite) set of
input-conditioned transition matrices over the latent state set $\MxSSet$: $\{
\T^{(\MxSSet \to \MxSSet | x)} \}_{x \in \mathcal{X}}$, where $\T^{(\MxSSet \to
\MxSSet | x)}_{i,j} = \Pr(\MxSt_{t} = r^j | \MxSt_{t-1} = r^i , X_t = x)$.
Probabilities regarding actual state paths can be obtained from the
latent-state-to-state transition dynamic together with the observable-state
projectors, which we now define.

We denote distributions over the latent states as bold Greek symbols, such as
$\boldsymbol{\mu}$. As in the main text, it is convenient to cast
$\boldsymbol{\mu}$ as a row-vector, in which case it appears as the bra
$\bra{\boldsymbol{\mu}}$. The distribution over latent states $\MxSSet$ implies
a distinct distribution over observable states $\SSet$. A sequence of driving
inputs updates the distribution: $\boldsymbol{\mu}_{t+n}(\boldsymbol{\mu}_{t},
x_{t:t+n})$. In particular:
\begin{align*}
\bra{\boldsymbol{\mu}_{t+n}} 
&= \bra{\boldsymbol{\mu}_{t}} \T^{(\MxSSet \to \MxSSet | x_{t:t+n} )} \\
&= \bra{\boldsymbol{\mu}_{t}} \T^{(\MxSSet \to \MxSSet | x_{t} )} \T^{(\MxSSet \to \MxSSet | x_{t+1} )} \cdots \T^{(\MxSSet \to \MxSSet | x_{t+n-1} )} ~.
\end{align*}
(Recall that time indexing is denoted by subscript ranges $n:m$ that are left-inclusive and right-exclusive.)
An infinite driving history $\vx$ induces a distribution $\boldsymbol{\mu}(\vx)$ over the state space, 
and $\boldsymbol{\pi_x}$ is the specific distribution induced by tireless
repetition of the single environmental drive $x$. 
This is the so-called ``equilibrium distribution'' associated with
equilibrating with the environmental drive $x$.
Explicitly:
\begin{align*}
\bra{\boldsymbol{\pi_x}} = \lim_{n \to \infty} \bra{\boldsymbol{\mu}_0} \left( \T^{(\MxSSet \to \MxSSet | x)} \right)^n ~.
\end{align*}
Usefully, $\boldsymbol{\pi_x}$ can also be found as the left eigenvector of
$\T^{(\MxSSet \to \MxSSet | x)}$ associated with the eigenvalue of unity:
\begin{align}
\bra{\boldsymbol{\pi_x}} = \bra{\boldsymbol{\pi_x}} \T^{(\MxSSet \to \MxSSet | x)} ~.
\label{eq:eigenpix2}
\end{align}
The canonical equilibrium probabilities are this vector's projection onto
observable states: $\pi_x(s) = \braket{\boldsymbol{\pi_{x}} | s}$, where
$\ket{s} = \ket{\delta_{s,f(r)}}$ has a vector-representation in the
latent-state basis with elements of all $0$s except $1$s where the latent state
maps to the observable state $s$.

Assuming latent-state-to-state transitions are Markovian allows the
distribution $\boldsymbol{\mu}$ over these latent states to summarize the
causal relevance of the entire driving history. 

\subsubsection{Implications}

A semi-infinite history induces a particular distribution over system latent
states and implies another particular distribution over its observable states.
This can be usefully recast in terms of the ``start'' (or initial) distribution
$\boldsymbol{\mu}_0$ induced by the path $x_{-\infty:1}$ and the driving
history $x_{1:t+1}$ since then, giving the entropy of the induced state
distribution:
\begin{align*}
h^{(s | \boldsymbol{\mu}_0 , x_{1:t+1} ) } 
  & = - \ln \Pr(\St_t = s | \boldsymbol{\mu}_0 , x_{1:t+1}) \\
  & = - \ln \bra{\boldsymbol{\mu}_0} \T^{(\MxSSet \to \MxSSet | x_{1:t+1})}  \ket{s}
  ~.
\end{align*}
Or, employing the new distribution and the driving history since then, the path
entropy (functional of state and driving history) can be expressed simply in
terms of the \emph{current} distribution over latent states and the candidate
observable state $s$:
\begin{align*}
h^{(s | \boldsymbol{\mu} ) } 
  & = - \ln \Pr(\St_t = s | \MxSt_{t} \sim \boldsymbol{\mu} ) \\
  & = - \ln \braket{\boldsymbol{\mu} | s}
  ~.
\end{align*}

Averaging the path-conditional state entropy over observable states again gives
a genuine input-conditioned Shannon state entropy:
\begin{align*}
\braket{h^{(s_t | \pastxt)} }_{\Pr(s_t | \pastxt)} 
  & =  \H [\St_t | \pastXt = \pastxt ]
  ~.
\end{align*}
It is again easy to show that the state-averaged path entropy $\kB \H[\St_t |
\pastxt ]$ is an extension of the system's steady-state nonequilibrium entropy.
In steady-state, the state-averaged path entropy reduces to:
\begin{align*}
\kB \H [\St_t | \pastXt = \dots xxx ]
  & =  - \kB \H[\St_t | \MxSt_t \sim \boldsymbol{\pi}_x ] \\ 
  & = - \kB \sum_{s \in \SSet} \pi_x(s) \ln \pi_x(s) \\
  &= S_\text{ss}(x)
  ~.
\end{align*}

The \emph{nonsteady-state addition to free energy} is:
\begin{align*}
\beta^{-1} \gamma(s | \boldsymbol{\mu} , x)
  \equiv \beta^{-1}
  \ln \frac{\Pr(\St_t = s | \MxSt_{t-1} \sim \boldsymbol{\mu}, X_{t} = x )}{ \pi_{x}(s ) }
  ~.
\end{align*}
Averaging over observable states this becomes the relative entropy:
\begin{align*}
\braket{\gamma(s | \boldsymbol{\mu} , x) } 
=  D_\text{KL} \left[ \Pr(\St_t | \MxSt_{t-1} \sim \boldsymbol{\mu} ,  X_{t} = x ) || \boldsymbol{\pi}_{x} \right]
  ~,
\end{align*}
which is always nonnegative.

Using this setup and decomposing:
\begin{align*}
\frac{\Pr(\St_{0:N} = s^0 \sF  | \MxSt_{-1} \sim \StartF , X_{0:N} = x^0 \xF )}{\Pr(\St_{0:N} = s^{N-1} \sR | \MxSt_{-1} \sim \StartR , X_{0:N} = x^N \xR )}  
\end{align*}
in analogy with Eq.~\eqref{eq:GenDetailedFT}, it is straightforward to extend
the remaining results of the main body to the setting in which observed states
are functions of a Markov chain. Notably, the path dependencies pick up new
contributions from non-Markovity. Also, knowledge of distributions over latent
states provides a thermodynamic advantage to Maxwellian Demons.

\section{Integral fluctuation theorems with auxiliary variables}
\label{sec:FTappendix2}

Recall that we quantify how much the auxiliary variable independently informs
the state sequence via the nonaveraged conditional mutual information:
\begin{align*}
i[\vsF; \vyF | \vxF , \StartF ]
  & \equiv \ln
  \frac{\Pr(\vsF , \vyF | \vxF , \StartF )}
  { \Pr( \vyF | \vxF , \StartF ) \Pr( \vsF | \vxF , \StartF ) }
  \\
  & = \ln 
  \frac{\Pr(\vsF , \vyF , \vxF | \StartF )}
  { \Pr( \vyF , \vxF , \StartF ) \Pr( \vsF | \vxF , \StartF ) }
  ~.
\end{align*}
Note that averaging over the input, state, and auxiliary sequences gives the
familiar conditional mutual information:
\begin{align*}
\I[\St_{0:N}; Y_{0:N} & | X_{0:N} , \StartF ] \\
  & = \left\langle
  i[\vsF ; \vyF | \vxF , \StartF ]
  \right\rangle_{\Pr(x_{0:N} , s_{0:N} , y_{0:N} | \StartF )}
  .
\end{align*}
(Averaging over distributions is the same as being given the distribution,
since the distribution over distributions is assumed to be peaked at
$\StartF$.)

Noting that:
\begin{align*}
& e^{\beta W_\text{diss} + i(\vsF ; \vyF | \vxF , \StartF ) + \Psi } \\
  & \qquad\qquad  = e^{\eeep + i(\vsF ; \vyF | \vxF , \StartF ) + \Psi + (\gamma_\text{F} - \gamma_\text{R} )} \\
  & \qquad\qquad  = \frac{\Pr(\vsF , \vyF , \vxF | \StartF )}
  { \Pr( \vyF , \vxF | \StartF ) \Pr( s^{N-1} \sR | \xR x^0 , \StartR ) } \\
  & \qquad\qquad  = \frac{\Pr(\vsF , \vyF , \vxF | \StartF )}
  { \Pr( \vyF , \vxF | \StartF ) \Pr( \vsR | \vxR , \StartR ) }
  ~,
\end{align*}
where $\StartR = \boldsymbol{\mu}(\StartF, \vx)$, we have the integral
fluctuation theorem (IFT):
\begin{align*}
& \left\langle
  e^{-\beta  W_\text{diss} - i(\vsF; \vyF | \vxF , \StartF ) - \Psi }
  \right\rangle_{\Pr(\vsF , \vyF , \vxF | \StartF )} \\
  & \qquad = \sum_{\vxF, \vsF, \vyF} \Pr(\vsF , \vyF , \vxF | \StartF )
  \frac{\Pr(\vyF , \vxF | \StartF ) \Pr( \vsR | \vxR , \StartR )}
  {\Pr(\vsF , \vyF , \vxF | \StartF )} \\
  & \qquad = \sum_{\vxF, \vsF, \vyF} \Pr( \vyF , \vxF | \StartF )
  \Pr( \vsR | \vxR , \StartF ) \\
  & \qquad = \sum_{\vxF, \vyF} \Pr( \vyF , \vxF | \StartF )
  \sum_{\vsR} \Pr( \vsR | \vxR , \StartR ) \\
  & \qquad = \sum_{\vxF, \vyF} \Pr( \vyF , \vxF | \StartF ) \\
  & \qquad = 1
  ~.
\end{align*}
Notably, this relation holds arbitrarily far from equilibrium and allows for
the starting and ending distributions to both be nonsteady-state. It is
tempting to conclude that the revised Second Law of Thermodynamics should read:
\begin{align}
\left\langle W_\text{diss} \right\rangle  
\ge -  \kB T \, \I[\vec{\St}; \vec{Y} | \vec{X} , \StartF ]
  - \left\langle \Qhk \right\rangle
  ~,
\label{eq:RevisedSecondLaw}
\end{align}
which includes the effects of both irreversibility and conditional mutual
information between state-sequence and auxiliary sequence, given
input-sequence. However, we expect that $\left\langle \Qhk \right\rangle > 0$,
so Eq. (\ref{eq:RevisedSecondLaw}) is not the strongest bound derivable.
Dropping $\Psi$ from the IFT still yields a true equality, but the derivation
runs differently since it depends on the normalization of the conjugate
dynamic.  Although IFTs with $\Psi$ may be useful for other reasons, it is the
non-$\Psi$ IFTs that seems to yield the tighter bound for the revised Second
Laws of information thermodynamics without detailed balance.

\bibliography{thermobib}
 
\end{document}